\documentclass[preprint,number,12pt]{elsarticle}

\usepackage{graphicx}
\usepackage{mathptmx}      
\usepackage{amssymb}
\biboptions{comma}
 
\journal{Astroparticle Physics}

\begin{document}

\begin{frontmatter}

\title{High energy gamma-ray emission from compact Galactic sources in the context of observations with the next generation Cherenkov Telescope Arrays}


\author{W\l odek Bednarek}
\address{Department of Astrophysics, The University of Lodz, ul. Pomorska 149/153, 90-236 Lodz, Poland}

\begin{abstract}
The observational progress in the $\gamma$-ray astronomy in the last few years has led to the discovery of more than a thousand sources at GeV energies and more than a hundred sources at TeV energies. A few different classes of compact objects in the Galaxy have been established. They show many unexpected features at high energies the physics of which remains mainly unknown. At present it is clear that detailed investigation of these new phenomena can be performed only with the technical equipment which offer an order of magnitude better sensitivity, and a few times better energy, angular and time resolution in the broad energy range staring from a few tens of GeV up to a few hundreds TeV. Such facilities can be realized by the next generation of instruments such as the planned Cherenkov Telescope Array (CTA).
   
The aim of this report is to summarize up to date observational results on the compact galactic sources in the GeV-TeV $\gamma$-ray energy range, discuss their theoretical implications, and indicate which hypothesis considered at present might be verified with the next generation of telescopes. We point out which of the observational features of the $\gamma$-ray sources are important to investigate with  special care with the planned CTA in order to put a new light on physical processes involved. Their knowledge should finally allow us to answer the question on the origin of energetic particles in our Galaxy.
\end{abstract}


\end{frontmatter}


\section{Introduction}
\label{}

It is widely believed that relativistic, charged particles are accelerated in the Galaxy up to at least $\sim$10$^{17}$ eV. 
Such particles are confined within the Galactic disk since their Larmor radius is smaller than the thickness of the Galactic disk.
However, which types of the observed sources are responsible for the acceleration processes (maybe a few ?) is at present not quite clear. The best candidates seem to be sources containing compact objects such as rotating neutron stars (NSs), solar mass black holes, massive stars, or maybe also White Dwarfs (WDs). They can either form separate systems which energize their surrounding (e.g. Pulsar Wind Nebulae) or occur close to concentrations of matter within the Galaxy (Globular and Open Clusters) or be companions of normal stars (binary systems).  

During the last few years our knowledge on the compact sources of cosmic accelerators in the Galaxy has experienced a major breakthrough. Thanks to the development of a new satellite and on-ground telescopes, 
a few hundred of discrete $\gamma$-ray sources at energies above 100 MeV have been discovered in the Galactic plane by the Large Area Telescope on board of Fermi satellite (Fermi-LAT)~\cite{nol12}, and a few tens of sources above $\sim$100 GeV have been discovered by the Cherenkov telescopes (see for the reviews, e.g.~\cite{hh08,ah08a,buc08}.
Some of these sources have showed unexpected emission features 
which put new light on the particle acceleration and 
radiation processes occurring in them.  

One of the purposes of this paper is to present up-to-date knowledge (both observational and theoretical) on the emission features of at present very well established $\gamma$-ray sources such as: different types of pulsars, Pulsar Wind Nebulae (PWNe), Globular Clusters (GCs) and 
binary systems containing either two massive stars, a WD, a pulsar, or a solar mass black hole. We also mention other potential compact $\gamma$-ray sources in the Galaxy such as: isolated massive stars and WDs, or accreting NSs and WDs. They are expected to emit $\gamma$-rays based on the theoretical analysis of the conditions within them which seem to be suitable for acceleration of particles and subsequent production of high energy radiation. 
We will consider how our knowledge of high energy processes in the Galactic sources can profit from the construction of the next generation telescopes able to observe $\gamma$-rays with energies above $\sim$30 GeV (such as the Cherenkov Telescope Array - CTA, the High Altitude Water Cherenkov Experiment - HAWC, or the Water Cherenkov Detector Array (WCDA) of the Large Altitude Air Shower Observatory - LHAASO). For example, CTA is planning to reach 50 hr sensitivity ($5\sigma$ detection) on the level of $1\%$ Crab flux units (CU) at $\sim$30 GeV and above $\sim$30 TeV, and $0.1\%$ CU in the range from 100 GeV to 3 TeV~\cite{ac11}. This will be about an order of magnitude better than achieved with the present telescope systems. The angular resolution of CTA is expected to be at least a factor of 2-3 better than that of the presently operating HESS Array.  
On the other hand, HAWC will provide all sky monitoring with 15 times better sensitivity than the Milagro Observatory (see HAWC Gamma-ray Observatory web page: http://hawc.umd.edu). It is expected that instruments with such sensitivities will finally allow us: to answer the question about the acceleration sites of particles with energies at most $\sim$10$^{15}$ eV, to conclude about their  acceleration mechanism, and provide constraints on the physics of these high energy sources. These new instruments are also expected to discover new types of sources, maybe some of them mentioned in this review.

\section{Isolated Rotating Neutron Stars}

Isolated rotating neutron stars (pulsars) produce winds and radiation that prevents accretion of matter onto their surfaces. They are at present one of the best known and defined astrophysical objects observed in the broad energy range, from radio up to $\sim$100 GeV. Their parameters can be precisely measured (period, period derivative) making it possible to estimate their basic physical quantities in terms of the rotating dipole model such as: the age, the surface magnetic field strength, the rotational energy loss rate. These characteristics allow us at present to distinguish three basic types of pulsars: (1) classical pulsars (CPs) with the surface magnetic field close to a few $\sim$10$^{12}$ G and rotational periods between several milliseconds and a few seconds; (2) millisecond pulsars (MSPs) with the surface magnetic fields close to a few $10^8$ G and periods of a few milliseconds; and (3) anomalous X-ray pulsars (AXPs) and soft $\gamma$-ray repeaters (SGRs) interpreted in terms of the magnetar model in which a slowly rotating neutron star (period of a few to several seconds) has a superstrong surface magnetic field of the order of $\sim$10$^{14}$ G. These specific classes of pulsars are not well separated. Objects with intermediate parameters are also observed.
Their existence can tell us a lot about the origin of specific classes, especially in the case of the nature of AXPs and SGRs since their magnetar interpretation has sometimes been questioned. Recent discoveries of the pulsed  $\gamma$-ray emission at energies $>$100 MeV from the high field magnetic radio pulsars, such as PSR B1509-28 \cite{abdo10}, the pulsar in CTA 1~\cite{ab08} and PSR J1119-6127~\cite{par11}, with the features similar to CPs have strengthened the magnetar hypothesis. 

Due to different parameters of these types of pulsars we consider below their possible high energy processes separately.
For a recent review of the basic properties of these pulsars see~\cite{lor11}.

\subsection{Classical pulsars}

With the launch of the Fermi and AGILE satellites, the number of known classical $\gamma$-ray pulsars has increased by more than an order of magnitude from 6 (discovered by EGRET telescope) to almost $\sim$100~\cite{abdo10b} within only 2-3 years. The excellent sensitivity and timing of the Fermi-LAT allowed this collaboration to detect several pulsars in blind periodicity searches~\cite{ab09a} for the first time.

\subsubsection{New results on old $\gamma$-ray pulsars} 

Studies of the strongest pulsars (detected already by EGRET) showed a variety of interesting details. For example the two peak structure observed in the Crab  pulsar in the radio and $\gamma$-rays are slightly shifted by $\sim$0.3 ms~\cite{abdo10c}. The Crab pulsar light curve shows interpulse emission which has clearly harder spectrum than that coming from the peak regions. 
In the case of the Vela pulsar, the $\gamma$-ray baseline emission extends through $80\%$ of the light curve, showing an intriguing third peak, more evident at higher energies~\cite{ab09b,abdo10d}. The first peak in the light curve of Vela dominates at low energies, but it disappears completely at the highest energies (above a few GeV). A fit to the whole pulsed $\gamma$-ray spectrum by an exponential function does not reproduce correctly the data at the highest detected energies. The third famous pulsar,
Geminga, shows many similarities to the previous two pulsars, although up to now no radio emission has been observed from it. Geminga shows $\gamma$-ray emission at all phases which is consistent with $\gamma$-ray production at regions close to the light cylinder~\cite{abdo10e}. Another interesting pulsar
with quite strong surface magnetic field, PSR B1509-58, has been originally detected only by the COMPTEL telescope~\cite{kui99}. 
Now, the pulsed emission is also observed by the AGILE and Fermi-LAT at energies below $\sim$1 GeV~\cite{pel09,abdo10}. The $\gamma$-ray spectrum shows a sharp break at a few tens of MeV, a feature not observed up to now in any other pulsar.

The whole population of pulsars detected by the Fermi-LAT shows a variety of light curves. These light curves can be divided into three main groups: those with two peaks well separated (by $\sim$0.4-0.5), with two close peaks (separated by only $\sim$0.2) and those with a single broad peak~\cite{abdo10b}.
The differential spectra are well described by the power law functions with spectral indices close to 1.5 and exponential cut-offs at a few GeV. At present
it is difficult to conclude what is the conversion efficiency of the rotational energy loss rate to the $\gamma$-ray power due to the unknown 
geometry of the $\gamma$-ray emission and uncertain estimates of distances to the pulsars.

\subsubsection{Unexpected detections of the Crab above several GeV}

Pulsars have become interesting targets for the presently operating Cherenkov telescopes even before launching the Fermi-LAT detector. For the first time, the Crab pulsar was detected at energies above $\sim$25 GeV by the
MAGIC Collaboration, using the special pulsar trigger~\cite{al08a}. Some evidence of the signal at energies above $\sim$60 GeV has been also reported from this pulsar~\cite{al08}. The lack of a sharp cut-off in the pulsed Crab spectrum at a few GeV has been also confirmed by the Fermi-LAT observations up to a few tens of GeV~\cite{at09}.
Further observations with the MAGIC and VERITAS telescopes allowed the groups to measure the pulsed $\gamma$-ray spectrum from the Crab pulsar up to $\sim$400 GeV~\cite{al11a,ale11b,sai11}. This total pulsed $\gamma$-ray spectrum is well described by a simple power law extending from the last Fermi-LAT point (see Fig.~1). 
The $\gamma$-ray pulses become narrower for larger energies. The ratio of the height of the first pulse (P1) to that of the second pulse (P2) in the Crab decreases with energy.

\begin{figure}
\vskip 7.8truecm
\includegraphics{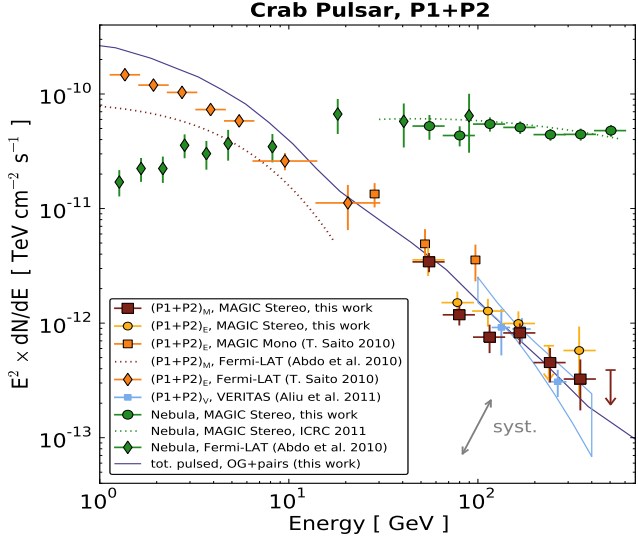}
\caption{The pulsed $\gamma$-ray spectrum of the Crab pulsar (both peaks together) observed with the Fermi, MAGIC and VERITAS telescopes.
Specific measurements are described in the legend within the figure
(from~\cite{ale11c}, reproduced by permission of the A\&A).}
\label{fig1}
\end{figure}

No other pulsed emission above a few tens of GeV has been reported up to by the modern telescope groups (e.g.~\cite{ah07a,al07,an10}. The pulsar spectra measured by Fermi-LAT seem to be consistent with exponential cut-offs at a few GeV~\cite{abdo10b}.

\subsubsection{Recent observations versus pulsar models} 

The $\gamma$-ray production in pulsars is interpreted in terms of a few general models locating the acceleration and $\gamma$-ray emission regions in different parts of the inner magnetosphere. Polar cap models~\cite{rs75,dh82} assume that acceleration occurs close to the polar regions of the NS surface. 
$\gamma$-rays are then produced by curvature radiation~\cite{dh82} or resonant/non-resonant Inverse Compton Scattering (ICS) of thermal radiation from the NS surface~\cite{sdm95}. The spectra are expected to show super-exponential cut-off at a few GeV due to the $\gamma$-ray absorption in a strong magnetic field. In a more recent version of this model, the so called pair starved polar cap (e.g.~\cite{ve09b}), particles are accelerated to large distances from the NS surface since the electric field is not saturated by inefficient production of $e^\pm$ pairs in the cascades.
In such a case, $\gamma$-rays can be produced at larger distances from the NS surface, not sufferring strong absorption in the magnetic field. Polar cap model can also be extended to the region close to the last open magnetic field lines up to the light cylinder where the curvature of magnetic field lines prevents production of the next generation of $e^\pm$ pairs along them~\cite{ar83,hm98}. A technical realization of this model has been successfully applied to a detailed modelling of a number of pulsar $\gamma$-ray light curves and spectra (the so-called two-pole caustic model~\cite{dr03}. In this model the $\gamma$-ray light curves are explained as the contributions from the regions of the two magnetic poles.

The acceleration of particles and $\gamma$-ray production 
can also occur in the outer pulsar magnetosphere in the region limited by the null surface and the light cylinder, according to the so called outer gap model~\cite{chr86a,chr86b,rom96}. 
The two versions of the model differ in the origin of the $e^\pm$ pairs which limit the extension of the outer gap. In the Cheng et al. model, $e^\pm$ pairs are
produced by the primary TeV $\gamma$-rays which collide with the synchrotron radiation from the outer gap. In the Romani model, the $e^\pm$ pairs are produced in collisions of primary curvature $\gamma$-rays with thermal X-rays from the NS surface.
In fact, the outer gap models predicted the existence of the low level, pulsed hard $\gamma$-ray spectrum extending up to TeV energies as a result of the comptonization of the synchrotron radiation by primary $e^\pm$ pairs. At present it is not clear
whether this component might be responsible for the sub-TeV pulsed emission from the Crab pulsar.
Detailed calculations of the expected $\gamma$-ray spectra and the light curves in this model are complicated since they require a three-dimensional treatment (see e.g.~\cite{tan08,ha08,hir08}. Depending on the parameters of the pulsars, $\gamma$-ray spectra are formed in the outer gap model either by synchrotron and curvature processes or, in the case of very young pulsars, also by the IC process~\cite{cz00,tak07}. The above considered models are based on simple pulsar magnetosphere structure (approximated by the vacuum dipole magnetic model). Recent numerical studies of rotating force-free pulsar magnetospheres enable one determine a magnetic field structure in the outer magnetosphere and, as a result, a more detailed calculations of $\gamma$-ray emission patterns (the so-called separatrix Layer model~\cite{bs10}). Based on these
calculations, it is concluded that the two-pole caustic and the outer gap models are not able to reproduce the observed pulse
structures in the $\gamma$-ray pulsar light curves.

Recent $\gamma$-ray observations made for the first time possible to put strong constraints on the $\gamma$-ray emission region within the pulsar magnetosphere, locating it at least a few stellar radii from the neutron star surface (closer in, the gamma-rays would be absorbed in the magnetic field and converted to $e^\pm$ pairs~\cite{bar04}). This actually eliminated the popular scenario for the $\gamma$-ray origin close to the stellar surface by the curvature radiation mechanism, the so-called polar cap model. However, these observations are still consistent with the pulsar models in which most of the $\gamma$-rays are produced far away from the neutron star surface, i.e. the slot gap and the outer gap models.

In the context of the recent discovery of the sub-TeV emission from the Crab pulsar, a production of $\gamma$-rays by the secondary cascade $e^\pm$ pairs in the IC scattering of the synchrotron UV-X-ray radiation from the gap was considered by Lyutikov~\cite{lyu11a} and the detailed calculations along such model were performed by Hirotani~\cite{hir11}, see also the comparison with the observations in~\cite{ale11b}. 

The pulsed sub-TeV component might also appear as a result of comptonization of the magnetospheric soft radiation by $e^\pm$ pairs in the relativistic pulsar wind above the light cylinder~\cite{ba00}. By tuning the distance of the emission region from the light cylinder and the Lorentz factor of the wind the emission was obtained in the sub-TeV region. The present observations allow us to put constraints on the wind formation region in the Crab nebula based on the last model. The pulsed $\gamma$-ray 
emission was also proposed to originate in the pulsar wind region in terms of the striped wind model~\cite{ki02,pet11}. The peaks of $\gamma$-ray emission
appear as a result of comptonization of the background radiation 
by electrons moving radially outside the light cylinder with the Doppler factor of the striped wind.

\subsubsection{Interesting problems to be studied with CTA}

The features of the Crab pulsar spectrum at sub-TeV energies were not predicted by any detailed model and appeared as a surprise for the pulsar community. Present observations suggest that 
the pulsed $\gamma$-ray spectrum from the Crab pulsar show a smooth transition between the GeV to sub-TeV energy range (above a break at a few GeV).
The question appears whether the sub-TeV pulsed emission means a  new 
component in the spectrum (another emission process as expected in some models?) or it is a continuation of the spectrum from the GeV energy range? 
Detailed observations of the transition between these two components with the 
next generation telescope systems (e.g. CTA) should solve this problem (i.e. whether two distinct components are present in the sub-TeV spectrum or where is the transition between these components?). It is also very important to establish the end of the TeV component in the pulsed Crab spectrum in order to constrain 
the maximum energies of emitting $e^\pm$ pairs and conclude about the exact mechanism of the $\gamma$-ray production. Are these $\gamma$-ray produced in the scattering process of the soft synchrotron radiation from the outer gap or the thermal component from the NS surface 
or maybe in the curvature process of particles accelerated close to the light cylinder?

Other pulsars discovered by the Fermi-LAT seem to have $\gamma$-ray spectra consistent with the exponential cut-offs as expected in the curvature radiation process. However, they were not observed with the Cherenkov telescopes up to now. One of the main tasks of CTA should be an extensive investigation of the large population of the $\gamma$-ray pulsars in order to determine which of them (maybe all?) show some level of sub-TeV emission? It is supposed, in terms of the outer gap model, that only very young pulsars with the Crab-like magnetospheres (and two well separated peaks) can produce sub-TeV emission. Observation of other classes of pulsars (Vela-type, Geminga-type or single pulsed) with CTA should significantly constrain the emission models. 

CTA should also be able to investigate in detail additional emission features (third peaks, shifts in localizations of peaks observed in different energies) in the pulsar light curves as observed in different energy ranges (starting from radio up to sub-TeV). This information will allow us to constrain relative location of different emission regions within the pulsar magnetosphere. Note that some theoreticians postulate the existence of a few emission regions (acceleration gaps) which might be responsible for the production of different parts of the pulsed spectrum. 
There is some evidence of the existence of the third peak in the pulsar's light curve, e.g. the case of Vela pulsar~\cite{abdo10d}, see also some features in the Crab light curve at sub-TeV energies~\cite{al11a,ale11b,sai11}.
Possible discovery of such other components by the CTA will put in trouble present general models for $\gamma$-ray emission. 

The localization of the acceleration and radiation regions within the pulsar magnetospheres (thanks to the precise 
sub-TeV measurements) will make possible a more precise modelling of the 
$e^\pm$ pair cascade processes and the $e^\pm$ pair content in the inner pulsar magnetosphere. 
Note that the present models are not even able to give satisfactory predictions of the numbers of $e^\pm$ pairs escaping through the pulsar light cylinder into the PWNe. The calculations of $e^\pm$ pair multiplicities in the inner pulsar magnetospheres~\cite{ha01} are at least an order of magnitude lower than their estimates based on the radiation output from their parent PWNe~\cite{buc10}. 
Sub-TeV $\gamma$-ray observations of the pulsed emission are also expected to constrain a possible role of ions in the pulsar magnetospheres (by limiting the regions of electron acceleration) which are expected to carry the currents necessary to support the pulsar acceleration and radiation mechanism. 

It will be also very interesting to study the stability of the sub-TeV pulsed component over long periods of time. Are the radiation processes responsible for the $\gamma$-ray emission related in some way to the pulsar glitches  or the giant radio pulses? The stability of sub-TeV emission and its possible relation to the variability observed in the radio  will require investigation of the $\gamma$-ray emission on short time scales. Therefore an instrument with an order of magnitude better sensitivity (like CTA) is very welcome.

\subsection{Millisecond pulsars}

Millisecond pulsars had not been widely expected to produce observable pulsed $\gamma$-ray emission  due to a much lower number of them observed in radio, their relatively low energy loss rates (much older objects,  with weak surface magnetic field), preferable occurrence in  binary systems, and theoretical predictions. Therefore, the discovery of a large number of MSPs by Fermi-LAT was rather a surprise. Before launching of the Fermi and AGILE telescopes~\cite{kui00} pulsed $\gamma$-rays were reported from only one MSP, PSR J0218+4232.

\subsubsection{Discovery of the MSP population}

At first, 8 MSPs were discovered by using radio ephemeris~\cite{ab09c}. Since MSPs are relatively close objects, many of them were expected to be found at high galactic latitudes. 
In fact, radio searches revealed several more MSPs in the Fermi-LAT error boxes of sources outside the Galactic disk. They show spectral properties typical for classical pulsars (i.e. the characteristic cut-off at a few GeV).
The $\gamma$-ray emission features of the MSPs resemble very much those observed in the case of CPs what suggests a common mechanism for $\gamma$-ray production. In spite of significant general differences 
between these populations of pulsars
(mentioned above), some other basic properties, e.g. the strength of the magnetic field close to the light cylinder, are very similar. This supports arguments that the emission region in MSPs is also located far away from the pulsar surface and the radiation process is similar. 

Some of the discovered MSPs are in binary systems containing underluminos companions (the so called Black Widow pulsars) or typical low mass companions (e.g.~\cite{cog11,kei11,ran11}.   
Many MSPs have also been discovered within Globular Clusters (GCs). Their cumulative emission is probably responsible for the observed $\gamma$-ray emission from several GCs (as discussed in more detail in Sect.~4). In fact, in the case of one MSP within GC, PSR J1824-2452 in M28, the pulsed $\gamma$-ray emission has been already reported~\cite{pel09}.

\subsubsection{$\gamma$-ray production models: expectations and observations}

In the past, MSPs were expected to be better sites for production of $\gamma$-rays with energies detectable by Cherenkov telescopes than classical pulsars due to significantly weaker surface magnetic fields and the resulting smaller $\gamma$-ray absorption~\cite{bul00,zc03,ha05}. The models discussed in those papers have predicted curvature $\gamma$-ray spectra extending up to a few tens of GeV, and 
also a much weaker separate IC component extending up to $\sim$1 TeV, produced by a comptonization of the thermal X-ray emission from the NS surface. The expectation of the curvature emission extending up to a few tens of GeV have not been confirmed by the Fermi-LAT observations. 

The calculations of the $\gamma$-ray spectra from MSPs, performed in terms of other models, showed breaks at lower energies 
(e.g.~\cite{ve05}). This is due to the farther location of  the $\gamma$-ray production region from the surface of the neutron star. These results are more consistent with the present Fermi-LAT observations. Recently, calculations of the $\gamma$-ray light curves observed from MSPs were made in terms of a few different models by Venter et al.~\cite{ve09a}. It is concluded that most MSPs require a location of the emission region in the outer magnetosphere, what is consistent with the prediction of the outer gap and slot gap models. This conclusion is similar to that reached from the analysis of the emission models of the CPs,
thus opening the possibility of a detection of sub-TeV emission also from the MSPs.

\subsubsection{Future CTA observations}

It is expected that the next generation Cherenkov telescopes will be able to establish possible sub-TeV $\gamma$-ray emission from MSPs. 
Such second $\gamma$-ray component in the pulsed spectrum, produced in the IC process, will be interesting to be investigated with the present and future Cherenkov telescopes.
Does this emission extends with a simple power law after the peak at a few GeV (as seems to be observed already in the Crab pulsar),
or is it a new separate component as expected in the earlier models for the MSP $\gamma$-ray emission? Note that due to the large differences in the location of the light cylinder radius in CPs and MSPs, the thermal radiation from the NS surface of the latter might be of much greater importance. In case of a positive detection, comparative studies of the sub-TeV $\gamma$-ray emission features of CPs and MSPs (such as $\gamma$-ray light curves, synchronization of the sub-TeV emission with emission in other energy ranges, cut-offs in the energy spectra),
will provide stringent constraints on the location of high energy processes in the pulsar magnetospheres with different magnetic fields and rotation periods. 

\subsection{Magnetars}

Magnetars (AXPs and SGRs) are rotating neutron stars with the surface magnetic field about two orders of magnitude larger than that observed in classical pulsars~\cite{dp92}. On the other hand, their rotational periods are much longer being in the range 2-16 s. It is expected that about $20\%$ of all neutron stars have been born as magnetars. Since their X-ray luminosity is clearly larger than the rotational energy loss rate, it has been proposed that these sources are powered by the decay of the super-strong magnetic fields in contrast to the MSPs and CPs powered by the rotational energy losses. Magnetars are relatively young objects, with the age comparable to the observed classical $\gamma$-ray pulsars, i.e. $10^3-10^5$ years.

\subsubsection{Expectations of pulsed $\gamma$-ray emission from magnetars}

The processes occurring within the magnetospheres of  
magnetars are expected to show general similarities to those occurring in the classical pulsars in spite of large differences in the surface magnetic fields. 
In fact, the values of the magnetic field strength close to the light cylinder in magnetars are of the same order as those in CPs. 
Therefore, the radiation models developed for CPs (especially their outer atmosphere versions) have been applied also for magnetars. The calculations performed in terms of the outer gap model~\cite{cz01,zc02} show that magnetars are likely sources of GeV $\gamma$-rays detectable by the Fermi-LAT telescope. These theoretical expectations were recently confirmed by independent calculations~\cite{ton11}. 

However, recent observations of a few magnetars do not confirm these predictions. The upper limits are clearly below predictions at least in the case of some magnetars (see~\cite{sas10,abdo10f}. This requires either a re-consideration of the applicability of the outer gap model for the magnetar magnetospheres or a consideration of another scenarios for the magnetar phenomenon. A way to understand the problem is the model of twisted magnetar atmospheres~\cite{bt09,bel09}.
In this model $e^\pm$ pairs are copiously produced already close to the surface of the neutron star as a result of magnetic $e^\pm$ pair production by $\gamma$-rays very close to their production place.
These pairs limit the available potentials within the magnetosphere to 
$\sim$10$^9$ V. In such case, only hard X-rays to soft $\gamma$-rays  (with very hard spectrum) can be produced by leptons which scatter resonantly thermal X-ray photons (in consistence with the observations~\cite{kui06,go06,en10}.

Note however, that there seems to be a continuum transition of properties between typical magnetars ($B > 10^{14}$ G) and CPs. Recently, GeV $\gamma$-ray emission was discovered from a strongly magnetised, rotation powered pulsar (PSR J1119-6127) with the surface magnetic field $4.1\times 10^{13}$ G, i.e. very close to the critical value $B_{\rm cr} = 4.4\times 10^{13}$ G when the electron cyclotron energy equals to its rest mass~\cite{par11}. In this case, quantum effects have to be taken into account when considering radiation processes close to the NS surface. This may significantly influence the possible development of cascades in the pulsar magnetosphere (see e.g.~\cite{hl06,su11}. However, the $\gamma$-ray emission from PSR J1119-6127 has features similar to those observed in CPs, again arguing for its origin in the outer magnetosphere where the magnetic field is relatively weak.

Halzen et al.~\cite{hal05} tried to estimate the fluxes of TeV $\gamma$-rays and neutrinos by simple extrapolation from the  hard-X-ray emission observed from magnetars. The $\gamma$-ray fluxes predicted in these calculations are clearly above the present observational limit reported by the VERITAS Collaboration for two magnetars~\cite{gue09}. Note that the magnetic field strength close to the light 
cylinder in magnetars, classical and millisecond pulsars is comparable. Therefore, it makes sense to expect pulsed $\gamma$-ray signals also from magnetars. This pulsed emission, if originated in the outer magnetosphere, could extend up to sub-TeV energies as recently discovered in the case of the Crab pulsar.  

\subsubsection{Can CTA detect $\gamma$-ray emission from magnetars?} 

Since the magnetic field strengths at the light cylinder in different types of rotating neutron stars are similar, it makes sense 
to search for pulsed sub-TeV emission also from magnetars.
The old arguments about the saturation of the electric field in the inner magnetosphere of the magnetar as a result of $\gamma$-ray absorption 
in a strong magnetic field may not be so convincing, as it was previously 
expected, due to the formation of the positronium stage by the $\gamma$-ray photons~\cite{su11}.

The rotational periods of magnetars were likely much shorter at birth than observed at present. As classical pulsars, they probably produce relativistic winds which terminate producing shock waves. Therefore, magnetars might also be surrounded by nebulae filled with relativistic particles as observed in the case of young pulsars. In fact, first extended TeV $\gamma$-ray source has been recently discovered in the direction of the well known magnetar SGR 1806-20~\cite{row11}.
A confirmation of the association of this TeV emission with magnetar by CTA would be of great importance for comparative studies of the properties of the relativistic winds from magnetars and classical pulsars and for the constraints of the multiplicity of $e^\pm$ pairs produced around NSs under very different conditions.
Note that the nebulae around magnetars are expected to be more extended due to a larger energy output at the early time after their formation. They should be characterised by weaker magnetic fields in comparison to the nebulae around classical pulsars. Therefore, they might resemble more the unidentified TeV sources which
are easier to discover in the TeV $\gamma$-rays than in the X-ray or radio observations.

\section{Pulsar Wind Nebulae}

The nebulae around energetic rotation powered pulsars have been suspected as sites of particle acceleration and $\gamma$-ray emission since their discovery (e.g.~\cite{gou65,rw69}. 
The high energy emission from these objects is expected to be produced by extremely energetic electrons interacting with the magnetic field 
and background radiation within the nebulae.
Up to now, several such nebulae have been reported to emit TeV $\gamma$-rays. A variety of their spectral and morphological features have been discovered. At present, however, they can not be investigated in enough detail due to the limited sensitivity, angular and energy resolution of the present telescopes.

\begin{figure}
\vskip 6.truecm
\includegraphics{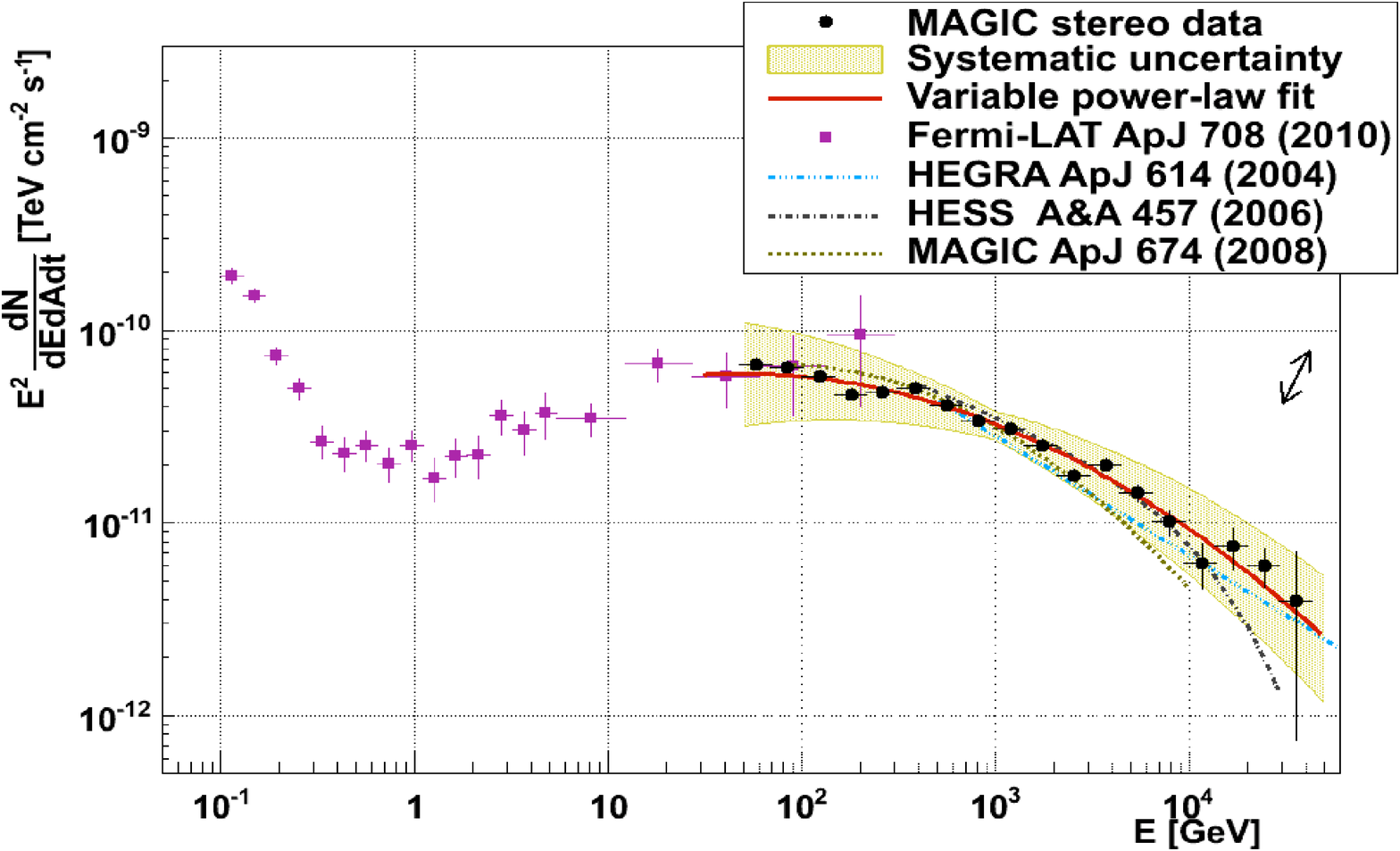}
\caption{Compilation of the TeV $\gamma$-ray measurements of the steady emission from the Crab Nebula (from~\cite{zan11}, reproduced by permission of the authors). Note the localization of the maximum in the spectrum (cumulative MAGIC and Fermi data) and the uncertainties in its shape above a few TeV.}
\label{fig2}
\end{figure}
\subsection{Observations of the Crab Nebula}

The Crab Nebula was the first object of this type which was 
established as a TeV $\gamma$-ray source about twenty years ago~\cite{wee89}.
Its multiwavelength spectrum is composed of two broad bumps wildly interpreted as the effect of the synchrotron and IC radiation of electrons with energies up to a few $10^{15}$ eV (e.g.~\cite{hj92,aa96}. The spectrum  showed an amazing stability within a few percents in most energy ranges. Therefore, the Crab Nebula has been considered as a standard source in the $\gamma$-ray astronomy~\cite{mey10}. However,
some variability at the end of the synchrotron spectrum (in the $\gamma$-ray range) on a time scale of a few months was mentioned based on the EGRET and COMPTEL observations~\cite{mun95,hj96}. Apart from these observations, the broadband spectrum of the Crab Nebula seemed to be well known and steady except the energy range above a few TeV where some differences might be noted resulting probably from the lack of statistics, see observations by the HEGRA~\cite{ah04}, HESS~\cite{ah06a}, and MAGIC telescopes~\cite{ale11d}. The Crab Nebula spectrum reported by Milagro in the energy range 1-100 TeV is generally consistent with the  previous HERGA measurement~\cite{abdo11d}. 
For a collection of these high energy observations see Fig.~2. 
Note that recently a long time scale, low level variability of the hard X-ray emission from the Crab nebula was reported (e.g.~\cite{wil11}. 

It came as a big surprise when at first the AGILE team~\cite{tav11} and next the Fermi-LAT team~\cite{abdo11b} reported the discovery of two large flares (flux increase by a factor of 4-6) lasting for several days at energies above $100$ MeV from the Crab Nebula (see Fig.~3). More detailed analysis of the September 2010 flare shows separate spikes on a half-day time scale~\cite{bal11}. Even more drastic flare from
the Crab Nebula was observed in April 2011 (see Fig.~4, the details reported in~\cite{bue11,st11}). During this flare the emission increased by a factor of $\sim$30 on an hour time scale.  The averaged spectrum of this flare has been well described by a power law with the spectral index 1.6 and the exponential break at $\sim$580 MeV~\cite{bue11}. Moreover, no accompanying variability in the other energy ranges have been observed, except the report on a simultaneous enhancement of the signal at $\sim$1 TeV by the ARGO-YBJ Collaboration~\cite{ai10}, not confirmed by MAGIC~\cite{mar10}, VERITAS~\cite{ong10} or MILAGRO~\cite{br11}.
\begin{figure}
\vskip 5truecm
\includegraphics{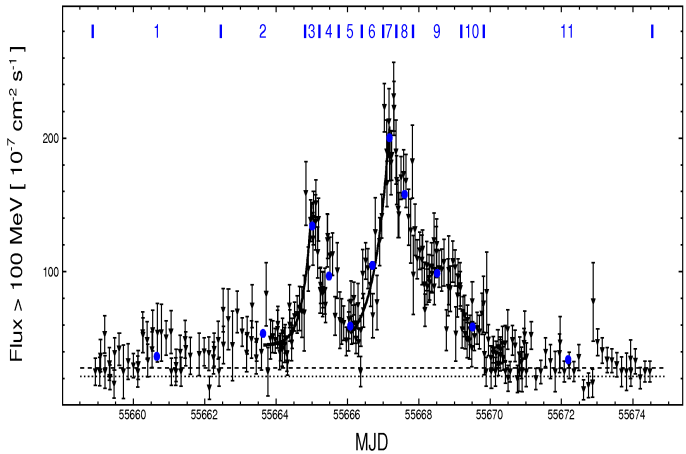}
\caption{The time structure of the flare observed in 2011 above $\sim$100 MeV~\cite{bue11}. Note the sub-day time scale variability of the $\gamma$-ray emission. (Reproduced by permission of the AAS)}
\label{fig3}
\end{figure}
\begin{figure}[t]
\vskip 10.3truecm
\includegraphics{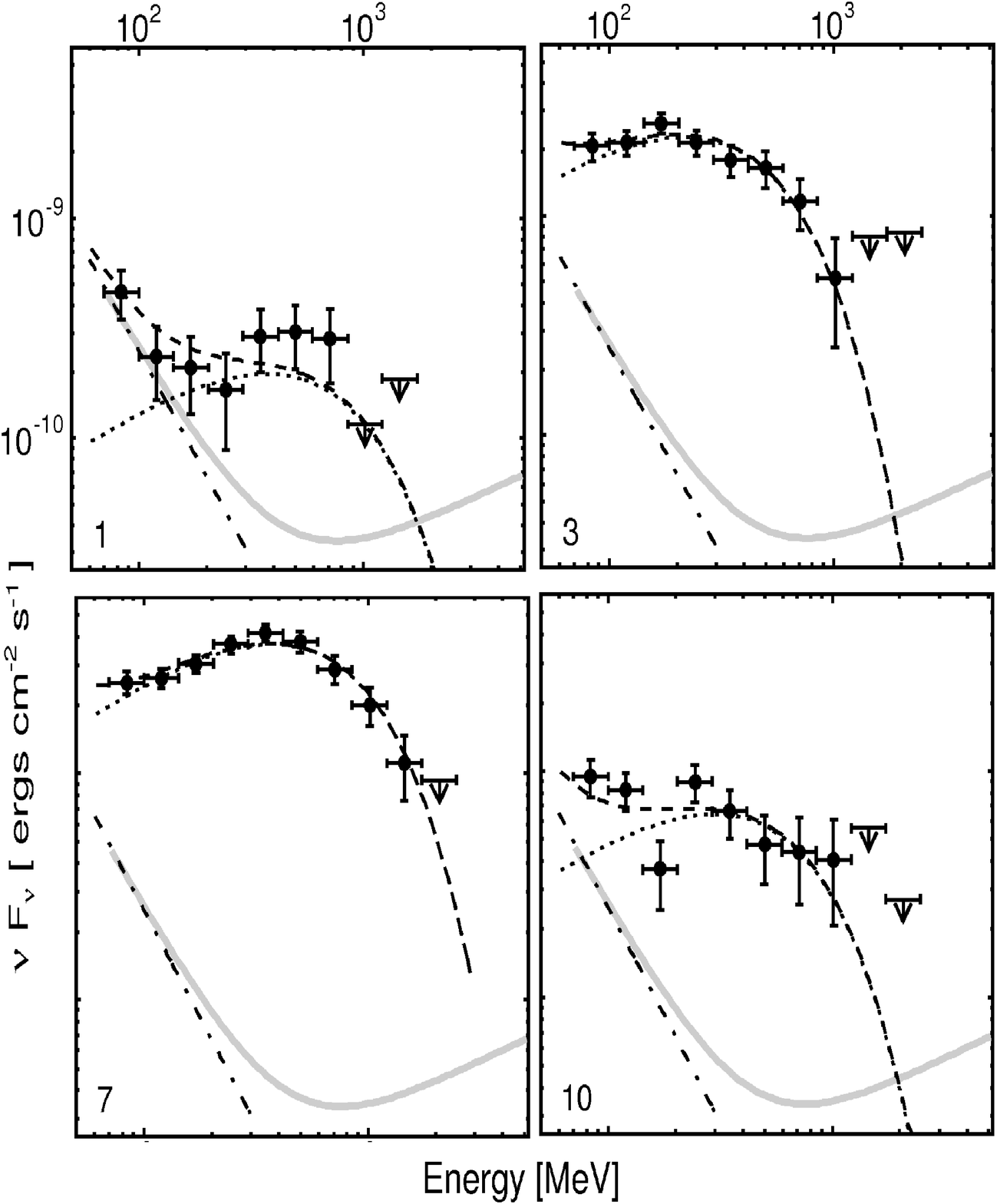}
\caption{The spectra of flaring $\gamma$-ray emission (see Fig.~3) from the Crab Nebula during the development of the flare in 2011 (from~\cite{bue11}). 
Note the behaviour of variable synchrotron spectrum observed at different moments of the flare and the maximum of the spectrum extending up to $\sim$1 GeV. (Reproduced by permission of the AAS)}
\label{fig4}
\end{figure}

\subsubsection{Interpretion of the variable $\gamma$-ray emission from the Crab Nebula} 

These observations pose serious problems for the popular mechanism of particle acceleration at the pulsar wind shock since they seem to violate the constraints on the maximum energy of accelerated particles, 
obtained by assuming diffusive shock acceleration or any mechanism connected  to acceleration in a magnetic field on a timescale longer than the gyration timescale.
Note that this limitation can be overcome in the case  of the magnetosonic wave interaction with the termination shock~\cite{lyu11b}. Moreover, the emission region is expected to move relativistically towards the observer (as considered in~\cite{bi11}) and/or the acceleration of particles can occur in other acceleration mechanism such as e.g., reconnection of the magnetic field in the pulsar wind~\cite{bi11,uz11} or in the large scale electric field generated by the rotating pulsar~\cite{abdo11b}. In another model, a variable synchrotron emission is proposed to be produced due to stochastic variations of the magnetic field in the reconnection region~\cite{byk11}.
The acceleration site (the place of the flare origin) is also enigmatic. It has to be very compact, i.e. much smaller than the travel time of light across the shock region. It has to be identified with a small region within the pulsar wind, e.g. inner knot? (see~\cite{kom11}. 

This surprising phenomenon will certainly reach much attention of the CTA, HAWC and other Cherenkov telescope projects. The Crab Nebula monitoring at multi-TeV energies by the water Cherenkov detectors seems to be of great importance since the flaring synchrotron component is expected to be produced by the $\sim$PeV leptons which might eventually also produce flaring Inverse Compton emission at multi-TeV energies. This can only occur when some specific conditions are fulfilled, i.e. leptons can cool on the ICS on the time scale of the flare.
This process can become efficient in the case of a significant Doppler boosting of the acceleration region of leptons~\cite{bi11}.
On the other hand, a possible discovery of variable multi-TeV emission from  the Crab Nebula with the next generation instruments would allow us to estimate the maximum energy of the accelerated leptons and thus constrain the emission region of the flaring component. Such variability would be in contradiction with the model~\cite{byk11}.

\subsubsection{Observations of other PWNe and their interpretation}

Observations of other PWNe at high energies show a variety of morphological and spectral features. In general, it is not so surprising since pulsars responsible for the PWNe are expected to be born with different initial parameters (period, surface magnetic field, velocity), sometimes in an inhomogeneous dense medium still present around them due to relatively short evolution time of their  progenitor stars. The parameters of a newly born pulsar determine the total energy content of the future nebula, the magnetic field inside it and also (likely) the maximum energy of the accelerated particles. On the other hand, the surrounding medium determines the size of the nebula, its morphology, and maybe also the dominant type of $\gamma$-ray radiation mechanisms (e.g. in some cases bremsstrahlung from electrons and pion decay from hadronic interactions). Recent observations, mainly in the radio, X-ray and TeV energy ranges, reveal this variety of features. However, the sensitivity and angular resolution of the present TeV $\gamma$-ray telescopes do not allow us to investigate of the details of the spectral and morphological features in a large population of objects
(only in exceptional objects energy dependent morphological studies have been performed). 
Therefore, robust conclusions about their physics can not be reached at present. Below we will review the most interesting features observed in some specific objects. 

A very extended TeV nebula ($\sim 0.5^{\rm o}\times 1.5^{\rm o}$), Vela X, has been discovered in the vicinity of the well known Vela pulsar. It is clearly shifted with respect to the pulsar location. The spectrum of the Vela X nebula is flat peaking at a few TeVs, atypical in comparison  to other PWNe~\cite{ah06b}. On the other hand, recently discovered GeV emission from the direction of Vela X~\cite{abdo10g,pel10} is much more extended, showing a spectrum with completely different features (see Fig.~5). The emission from a more compact TeV nebula (cocoon) was modelled assuming its hadronic~\cite{hor06} or leptonic origin~\cite{la08}. The complicated structure of the Vela nebula in different energy ranges was explained as due to the two populations of electrons injected by the pulsar at present and in the past~\cite{dj08}.

\begin{figure}
\vskip 8.truecm
\includegraphics{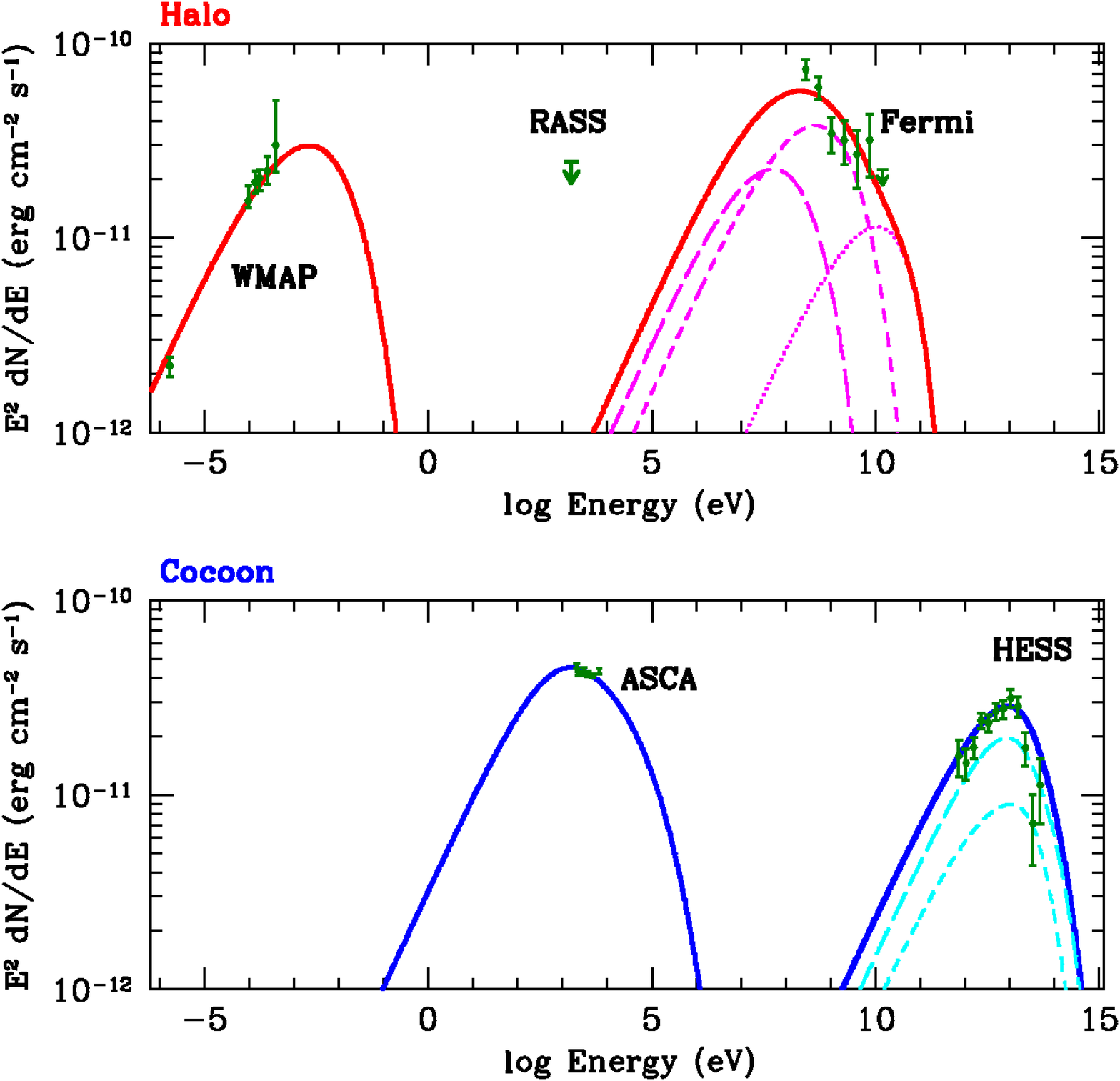}
\caption{Compilation of spectral measurements of different regions (upper panel - Halo and lower panel - Cocoon) in 
Vela-X pulsar wind nebula collected from different instruments (for details see~\cite{abdo10g}, reproduced by permission of the AAS). Note large differences in the spectral shape between the Cocoon (produced by fresh particles from the Vela pulsar) and the Halo (produced by old, lower energy particles).}
\label{fig5}
\end{figure}

Other PWNe, observed in the GeV and TeV energies, also show spectra with clear discontinuities suggesting an existence of two populations of particles,
e.g. HESS J1640-465~\cite{ah06c,sl10}. In the case of young objects this might be a result of the acceleration of particles within the PWNa itself and also on the supernova shell, sometimes visible in the so called composite PWNe. The two component spectra are not characteristic for all nebulae. For example, the nebula MSH 15-52 shows the GeV-TeV spectrum nicely described by the IC emission from a single population of electrons, and the radio to X-ray spectrum - by the synchrotron process from this same population of electrons~\cite{sa00,ah05a}.
As in the case of the Crab Nebula, the GeV spectra of other PWNe seem to be clearly harder than the TeV spectra (e.g. MSH 15-52 or PSR J1023-5746/HESS J1023-575~\cite{saz10,abr11b}. They are produced by electrons with lower energies characterised by long cooling time scale within the nebula. Their investigation should  provide information about the injection rate of electrons in the early age of the nebula. Therefore, precise measurements of the relative emission from nebulae in the GeV and TeV energy ranges should allow us to  estimate the total power in relativistic particles and so to constrain
the initial parameters of the parent pulsars (see e.g.~\cite{dj08b}. 

The morphological features of PWNe in the TeV and X-ray energies make possible to distinguish two groups which differ in the age. Young PWNe show compact (X-ray and TeV) nebulae with the pulsar more or less in the middle. On the other hand, older PWNe are extended with the pulsars not  located centrally. They are clearly more compact in X-rays than in 
TeV $\gamma$-rays. This feature has been expected since synchrotron X-ray  emission is produced by electrons with larger energies in the regions closer to the pulsar where the magnetic field is stronger.
On the other hand, the TeV emission comes from the electron IC scattering on the background radiation in the vicinity of the PWNa (such as the Cosmic Microwave Background (CMB) and the infrared-optical background in the Galactic disk).
These general features are consistent with the recent measurements of the structure of the PWNa, HESS J1825-137, showing a softening of the TeV spectrum with the increasing distance from the pulsar~\cite{ah06d}. 

The complex structure of the PWNe may be a consequence of its interaction with the asymmetric reverse shock of the supernova remnant~\cite{blo01}. This can result in a displacement of the nebula with respect to the pulsar as frequently observed in TeV energies. Also the high velocity pulsars can result in a formation of asymmetric PWNe. What influences the morphology of a specific nebula requires an investigation of its detailed structure in the $\gamma$-rays with a sensitive instrument and also measurements of the pulsar velocity and the PWNa surroundings. 
For example, in the case of the Vela PWNa, the asymmetric structure of the nebula with respect to the location of the pulsar is likely due to the interaction with the reverse shock since the pulsar moves in a completely different direction than the largest extent of the nebula~\cite{dod03}.

The structure of the PWNa may not only be determined by its background soft radiation content and the surrounding medium (including supernova remnant) but also it should depend on the geometrical structure of particle injection. It is well known from X-ray observations (e.g.~\cite{hes08} and MHD simulations (e.g. ~\cite{bk02,vol08} that the acceleration process of electrons is likely to occur highly anisotropically (e.g. the presence of the equatorial pulsar wind shock, jets, etc...). Probably we are in touch with such a case in the nebula MSH 15-52
around the pulsar PSR 1509-58, where the TeV nebula is aligned with the direction of the X-ray jet~\cite{ah05a}. 

Possible presence of dense clouds in the PWNa vicinity might result in
additional components in the GeV-TeV spectrum due to the interaction of electrons with matter (bremmstrahlung process) or possible interaction of hadrons (pion production). Such interacting PWNe might be good indicators of the acceleration of hadrons by the PWNe~\cite{bb08}. Detailed spectral and morphological studies with CTA and in other energy ranges  should answer the question whether such objects exist at all, what is the relative contribution to the high energy range from the PWNa, supernova shell, and particles which escaped from the PWNa and the supernova shell.

\subsubsection{How CTA can help to understand PWNe?}

Even in the case of the best studied objects, such as the Crab Nebula,
CTA with its angular and spectral resolution can 
answer some new enigmas. CTA is expected to contribute to the problem of the extent of the TeV source in the Crab Nebula.
Is it really a point like source (with a dimension corresponding to the 
pulsar wind shock) or is it comparable to the extent of the whole
nebula (2-3 arc min) or maybe even larger? Is it possible to constrain energy dependent dimension of the Crab Nebula emission?
Very important question is related to the variability of the Crab Nebula multiwavelenght spectrum. Can the whole spectrum above several GeV (Inverse Compton origin) vary on a long time scale as recently discovered in the X-rays? It would be very important to establish the spectral features at the highest energy end of the Crab Nebula spectrum (i.e. 
at $\sim$10 TeV). Is there any evidence of a contribution from an other radiation process (not IC) as postulated by some hadronic models
(e.g.~\cite{bp97,am03,bb03} ?
This part of the spectrum could also change on short time scale in accordance to the variable emission above $\sim$100 MeV since both are expected to be produced by electrons with energies of
a few hundred TeV. 
A determination of the end of IC $\gamma$-ray spectrum from the Crab Nebula by CTA would also allow us to constrain the magnetic field strength in the inner nebula since the location of the cut-off in the TeV region should also depend on the relative efficiency of the lepton cooling on the synchrotron and IC processes.

The very large angular dimensions of some nearby PWNe (even of the order of a few degrees) will require an instrument with a very large field of view. CTA, with a field of view even larger than that of the present HESS field of view and an order of magnitude better sensitivity, will make possible detailed studies of fine structures within PWNe. Note that old close nebulae are expected to contain pulsars which produce bow shocks due to their fast motion through the interstellar medium. On such bow shocks electrons might be accelerated up to the TeV energies. A few of such bow shocks have been observed up to now in the X-rays (e.g. 
around PSR J1747-2958,~\cite{gae04}) or around Geminga, e.g.~\cite{pav10}). The low level TeV emission from the bow shock, together with the large scale emission from the PWNa, produced by older electrons injected in the vicinity of this same pulsar, would enable us to make comparative studies of the two $\gamma$-ray objects created by the same pulsar. Detection of such PWNe with CTA would provide stringent constraints on evolution models of the PWNa and on the pulsar wind content.

Detailed investigation of energy dependent morphology of such composite PWNe with the future CTA will certainly allow to conclude on the location of the dominant particle acceleration site
within such nebulae.
In the case of such nebulae particles can be accelerated not only in the vicinity of the pulsar (pulsar wind shock, bow shock) but also by the supernova shocks (forward and reverse) appearing as a result of their interaction with the inhomogeneous surrounding medium.

It will be very interesting to study the spectral and morphological features of the Vela nebula from sub-TeV up to multi-TeV energies in order to find out whether the $\gamma$-ray emission has actually either a two component origin or it is due to a single component which changes continuously its properties with the distance from the center of the nebula. Only a telescope with the sensitivity and angular resolution of CTA can put new light on this problem. 

What factors are responsible for the variety of the spectral features observed in PWNe remains a mystery. This problem can be answered with accurate observations with CTA provided that it has good sensitivity already at $\sim$20-30 GeV and also above $\sim$30 TeV. The good sensitivity at large energies should allow us to constrain the maximum energies of accelerated particles. The information on the end of the electrons spectrum and on the end of synchrotron spectrum (expected in the hard X-ray - soft $\gamma$-ray energy range) will allow us to independently constrain the magnetic field strength within the PWNe. 

The radiation efficiencies of the PWNe are estimated in the range of a few up to $\sim 30\%$. They constrain the so-called sigma parameter
describing the ratio of energies injected into the nebula in the form of the magnetic field and relativistic particles. The knowledge of this value and its evolution with the  pulsar age is the basic factor limiting the multiplicity of pairs injected from the inner pulsar magnetosphere and the acceleration processes occurring in the pulsar wind region and/or the pulsar wind shock.  
Therefore, reliable measurement of the radiation efficiency in the large population of PWNe with CTA will be very important for constraining the pulsar models determining those of the PWNe.
 
CTA should also have a possibility to resolve fine structures (e.g. peaks?) in the TeV $\gamma$-ray spectrum which might be due to direct comptonization of the soft radiation by leptons in the relativistic pulsar wind~\cite{ba00}. Their discovery with CTA would indicate to the region of the relativistic wind formation and help to determine its Lorentz factor. We may wonder whether this Lorentz factor scale in a simple way with the basic parameters of pulsars and how it depends on the evolutionary stage of the PWNa.

\section{Globular Clusters}

Globular Clusters (GCs) are concentrations of $10^5-10^6$ old stars (with the age $>10$ Gyrs) contained within a spherical volume of a few pc. About 150 clusters, discovered up to now, create a spherical halo around the Galaxy with typical radius of $\sim$10 kpc~\cite{har96}. GCs also contain plenty of compact objects, millisecond pulsars (MSPs), Cataclysmic Variables (CVs) and Low Mass X-ray Binaries (LMXBs). Up to now, about $\sim$140 MSPs have been discovered in GCs~\cite{cr05,fre09}, and thousands of X-ray sources are observed. They are expected to belong to the class of CVs (e.g.~\cite{ver01}). It is argued that the numbers of these compact sources (MSPs, CVs, LMXBs) in a specific GC correlate with the so called encounter rate which is the combination of the core density and the core radius of GC~\cite{po03,po06,abdo10h}. 

\subsection{Gamma-ray observations}

Several Globular Clusters have been recently established as sources of GeV $\gamma$-rays in the observations with the Fermi-LAT telescope~\cite{ab09d,ko10}. They are characterised by differential spectral indices in the range $0.7\div 1.4$ and exponential cut-offs at $1.0\div 2.6$ GeV. The $\gamma$-ray power of various GCs differ by an order of magnitude. Energy spectra of GCs show features very similar to those recently observed in the population of the MSPs~\cite{abdo10h}. However these features may not be common for all GCs since some of the recently discovered objects show spectra extending above $\sim$10 GeV, without clear evidence of cut-offs~\cite{tam11a}. Therefore, the origin of $\gamma$-rays may not be only related to processes occurring within the inner magnetospheres of MSPs. 

The TeV emission has been searched from GCs already with the Whipple telescope (see the upper limits for M13 and M15~\cite{hall03,le03}). More stringent upper limits, at the level of present theoretical predictions, were obtained by modern Cherenkov telescopes for M13~\cite{an09}, M5, M13 and M 15~\cite{mc09}  and Tuc 47~\cite{ah09}.  
Only one TeV $\gamma$-ray source was discovered close to the GC, Ter 5~\cite{abr11a}. Ter 5 contains the largest number of the MSPs (33 objects). The observed $\gamma$-ray flux (1.5$\%$ of the Crab Units) is at the level expected from the models but the source is clearly shifted from the position of 
Ter 5. On the other hand, the position of the TeV $\gamma$-ray source is consistent with the location of the diffuse nonthermal X-ray source~\cite{eg10}  and the radio source~\cite{cla11}. However, why is the source shifted from the position of Ter 5 is at present unknown. 
The energy spectra observed from the direction of Ter 5 in the GeV and TeV energies can not be connected smoothly (see Fig.~6).
They seem to constitute two independent components. This suggests their  origin by different radiation mechanisms in different locations, e.g. GeV emission from the GC itself and 
the TeV emission from a completely different type of source, maybe 
even not directly related to the GC itself.

The $\gamma$-ray emission from GCs shows some interesting correlations with the GC content. For example, 
Abdo et al.~\cite{ab09c} and Hui et al.~\cite{hu09} find a clear correlation of the observed GeV $\gamma$-ray luminosity with the stellar encounter rate which determines the number of MSPs and Cataclysmic Variables within the specific GC. It is believed that the encounter rate determines the formation rate of compact binaries which are responsible for the origin of the MSPs inside GCs. The encounter rate can be estimated from the observed density of stars in the core of GC and the core radius~\cite{gen03}. 
Moreover, Hui et al.~\cite{hu09} report the positive correlation of the $\gamma$-ray luminosity with the metallicity [Fe/H] of the GC. They also find a tendency of the correlation of the $\gamma$-ray flux with the energy density of soft photon field at the GC location. 

The GCs identified with the Fermi-LAT $\gamma$-ray sources differ in their GeV $\gamma$-ray power by an order of magnitude~\cite{abdo10h}. Based on the observed $\gamma$-ray luminosity of a specific GC, the average energy loss rate of MSP (adopted for MSPs observed in Tuc 47~\cite{ab09c}, and the average spin-down to $\gamma$-ray luminosity conversion efficiency~\cite{ab09c,abdo10h} allows one to estimate the number of MSPs in specific GC.  The populations of MSPs in specific GCs derived by Abdo et al.~\cite{abdo10} are in good agreement with the estimates of the number of MSPs obtained with the help of other methods (see the estimates for Ter 5 in~\cite{fg00,ko10}).

\subsection{Theoretical predictions and confrontation with observations}

Since globular clusters contain many MSPs, it was expected that their cumulative GeV-TeV $\gamma$-ray emission might be above the sensitivity of the $\gamma$-ray telescopes. A few estimates of the $\gamma$-ray fluxes from MSPs in globular clusters were published already before the launching the Fermi Observatory. Based on the pair starved polar cap model Harding et al.~\cite{ha05} predicted 
the $\gamma$-ray fluxes from Tuc 47 near the EGRET upper limits. Venter \& de Jager~\cite{ve08}, applying the modern version of the polar cap model (general relativistic effects included), also predicted $\gamma$-ray spectra well above the sensitivity of the ${\it Fermi}$-LAT telescope. 
However, there are at present some doubts concerning the validity of the polar cap model as a likely scenario for the $\gamma$-ray emission from the pulsar inner magnetospheres (see Section~2).

\begin{figure*}[t]
\vskip 5.5truecm
\includegraphics{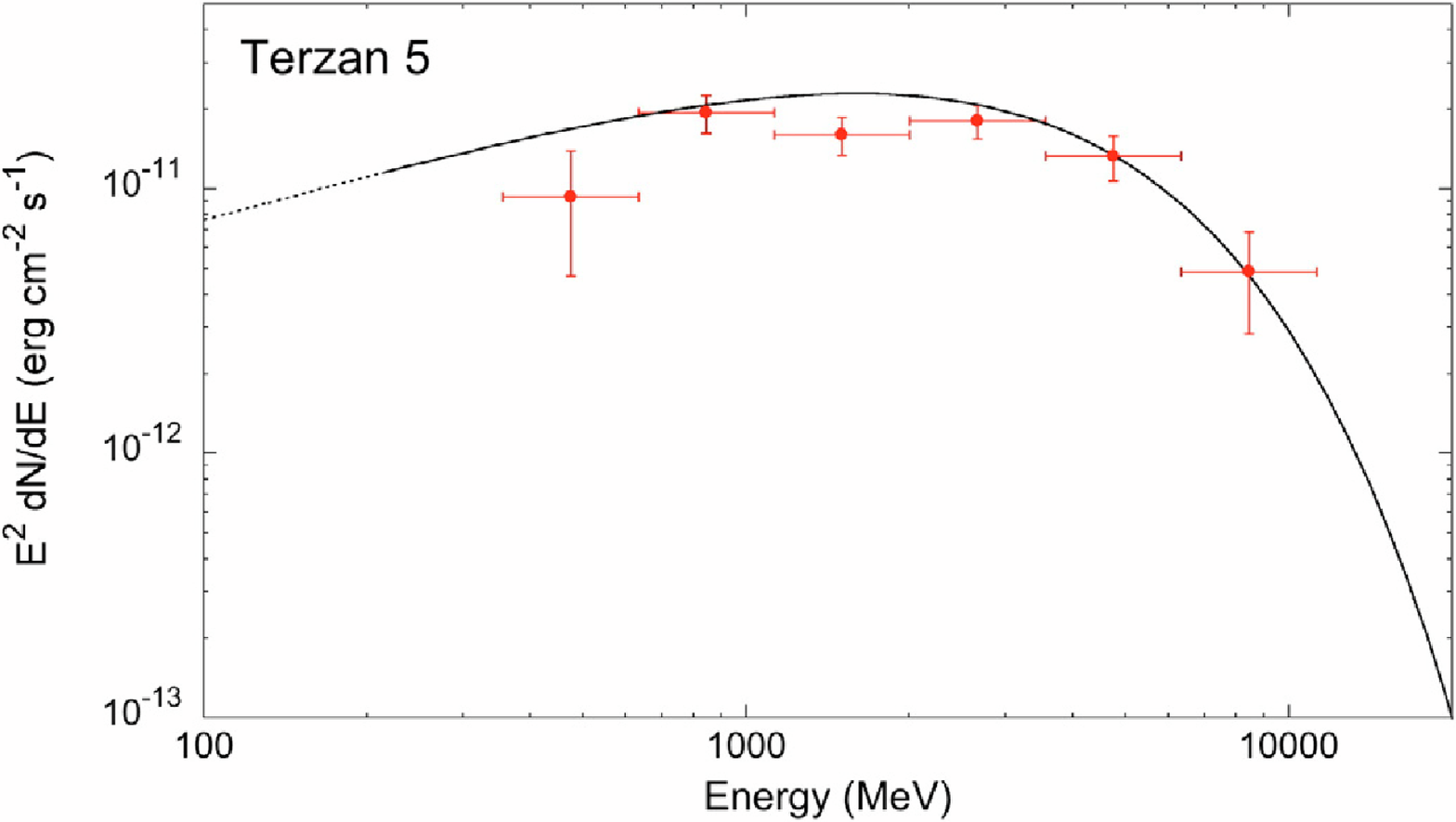}
\includegraphics{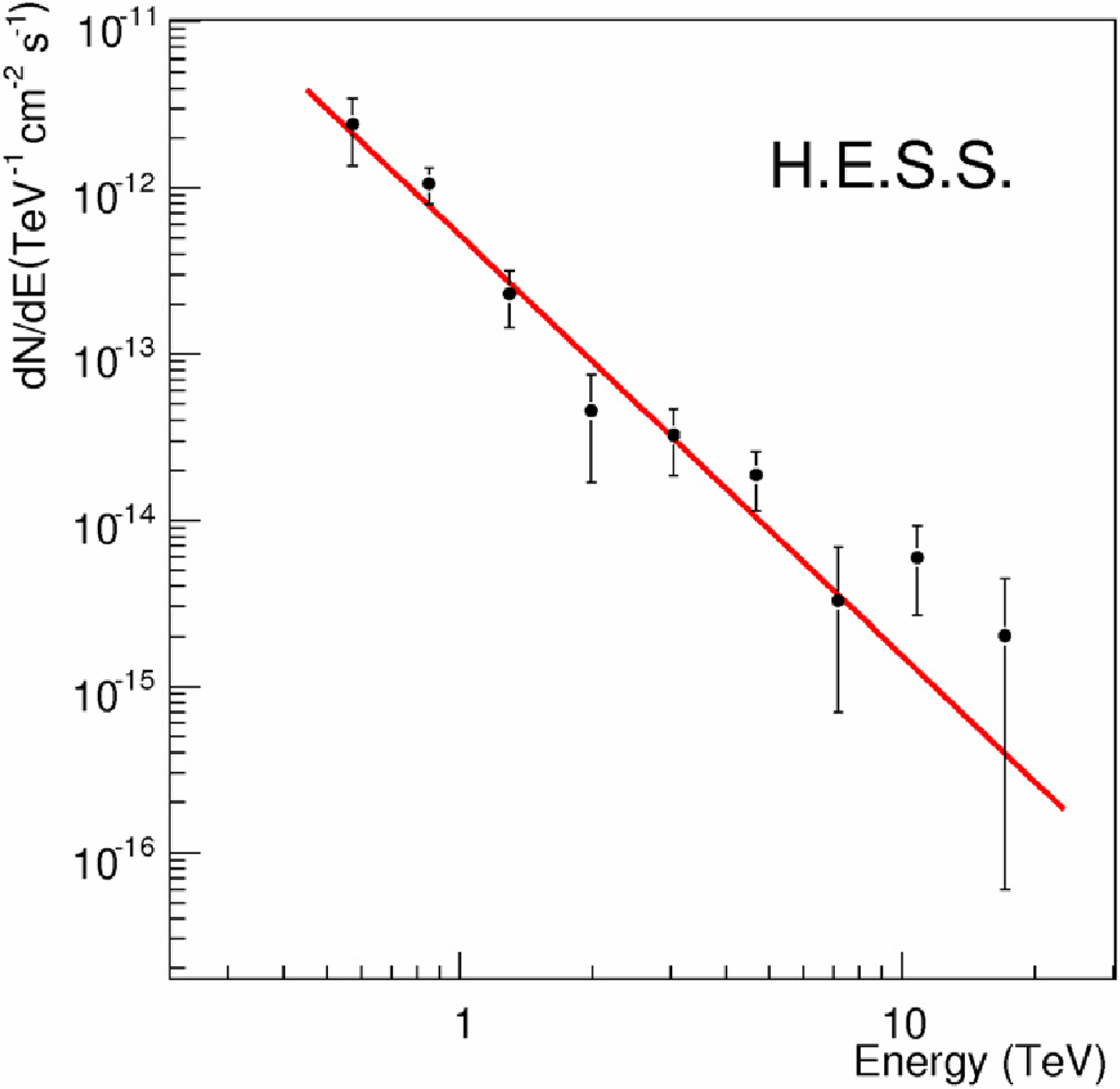}
\caption{The $\gamma$-ray spectra observed from the Globular Cluster Ter 5 in the GeV energy range (left figure from~\cite{abdo10h}, reproduced with permission of A\&A) and in the TeV energy range (right figure from~\cite{abr11a}, reproduced by permission of the A\&A). Note that the TeV component has likely different origin than the GeV component since the spectrum index, estimated from the two last points of the Fermi spectrum, is much steeper than the spectral index observed in the TeV energy range.}
\label{fig6}
\end{figure*}

The presence of a very dense radiation field within the clusters, produced by normal stars (2-3 orders of magnitude larger than the energy density of the CMB), suggests that relativistic particles injected by MSPs (and other compact sources) should efficiently produce TeV $\gamma$-rays in the Inverse Compton Scattering process. Such general scenario was considered for the first time by Bednarek \& Sitarek~\cite{bs07}. In this model, it is assumed that leptons are accelerated either in the millisecond pulsar winds or shocks produced by such winds (as a result of their collisions with the stellar winds or collisions of the two pulsar winds between themselves) or they are injected directly from the inner pulsar magnetospheres. 
These energetic leptons diffuse through the volume of the GC and interact at
the same time with its soft radiation, i.e with the optical photons from the stars and the CMB. 
Due to the lack of detailed knowledge about the spectral features of leptons injected by the MSPs, a few different models for electron injection are considered with the limitations expected from a comparison with the classical pulsar population.
The basic properties of the $\gamma$-ray spectra calculated in the case of mono-energetic injection of leptons into the soft radiation field dominated by soft photons
from the stellar population within the cluster and the CMB are shown in~\cite{bs07}. 
In terms of such a model, the $\gamma$-ray spectra were calculated for selected globular clusters assuming different spectra of injected leptons~\cite{bs07}.
A comparison of the predicted spectra with the observations of specific GCs allowed the authors to put the upper limits on the product of the number of MPSs within GC and their lepton injection rate.

Similar models were also considered in~\cite{ve09a,cheng10}. In the first paper it is assumed that leptons are injected from the inner magnetospheres of the millisecond pulsars
 through the light cylinder in the pulsar polar cap model~\cite{ve08}. 
In such a case, the injection rate of leptons is directly linked to the pulsed $\gamma$-ray emission from the MSPs observed in the GeV energies.
The TeV emission expected in this work is quite similar to that predicted by Bednarek \& Sitarek. On the other hand, Cheng et al.~\cite{cheng10} consider a production of $\gamma$-rays in the IC scattering on the galactic infrared and optical background by leptons in the pulsar winds. They predict an appearance of a relatively narrow second bump in the $\gamma$-ray spectrum from GCs at energies below $\sim$100 GeV. However, the abrupt cut-off in this spectrum above 100 GeV seems to be in contradiction with the observations of the TeV emission from Ter 5, provided that this emission comes from the GC itself.

It has also been suggested that a significant part of the TeV  emission observed from GCs may also come from other sources within the GC. For example, Bednarek~\cite{bed12} notes that a
typical GC should contain several thousands of WDs (with a significant fraction with the surface magnetic field of the order of $10^{8-9}$ G) which might be able to accelerate electrons to TeV energies (for more details see Section 9.3). 
On the other hand, Domainko~\cite{dom11} and Abramowski et al.~\cite{abr11a} propose that a recent type Ia supernova explosion or a short Gamma-Ray Burst can accelerate hadrons
which could produce $\gamma$-rays in the interactions with matter 
within or around the GC Ter 5.

\subsection{Future studies of GCs with CTA}

Sensitive observations of GCs in the sub-TeV energies (below $\sim$100 GeV) will be very important for a determination of a  relation between the observed spectra of the Fermi-LAT sources and the predicted TeV emission. Do GCs really emit TeV $\gamma$-rays produced by the leptons accelerated by the relativistic winds of the MSP population?

The next generation of Cherenkov telescopes should clarify the origin of the TeV source in the vicinity of Ter 5. Are there any morphological signatures of the relation of this source to Ter 5? This problem could be answered by the energy dependent morphological studies of the HESS source towards Ter 5.
For example, the closer geometrical connection between the HESS TeV source and the center of Ter 5 in the multi-TeV energies than in the sub-TeV energies will suggest that the relativistic particles originate within this GC. Detailed investigations of spectral and morphological features might also allow us to assign this source to the already known classes of the TeV sources (SNRs, PWNe, binary systems)?

Detection of sub-TeV pulsed emission from specific MSPs within the GCs would make possible to estimate relative contributions to observed GeV emission,
with respect to the total, of separate MSPs and that from the more extended component produced by relativistic leptons escaping from their magnetospheres or perhaps also from magnetized WDs.
This will allow to put additional constraints on the parameters of the supposed relativistic winds from the MSPs since they propagate in a well defined background radiation within the GCs.

Another problem to be answered is related to the types of sources are potentially able to produce $\gamma$-rays. Are the $\gamma$-ray spectra of GCs similar to each other or do they differ significantly forming separate classes? At present it can not be excluded that the contribution to the observed $\gamma$-ray emission from specific GCs is dominated by different objects expected to 
produce $\gamma$-rays such as: MSPs, magnetized WDs, Low Mass X-ray Binaries (LMXBs), type Ia supernova remnants or novae. There is also some evidence that the soft radiation content of several GCs can vary significantly~\cite{hu11}. It will be interesting to find out which of the two general types of GCs (centrally collapsed and extended) correlate in any way with their possible TeV emission. Only CTA with an order of magnitude better sensitivity can clarify these problems.

\section{Pulsars within binary systems}

For a long time X-ray binary systems have been suspected as sites of 
high energy processes which can lead to a production of $\gamma$-rays. However, the first claims of the TeV-PeV $\gamma$-ray detection from these sources were not confirmed by the more reliable observations with the Whipple Cherenkov telescope~\cite{wee92}. In the GeV energies, a few X-ray binary systems were detected in the large error boxes of the EGRET telescope (LS 5039~\cite{par00}, Cyg X-3~\cite{mor97}, LSI 61 303~\cite{tho95},  Cen X-3~\cite{ve97}). These associations were not conclusive since the $\gamma$-ray signals did not show any evidence of the expected modulation with the orbital period of the binary systems. 

\subsection{Recent observations in the GeV-TeV energy range}

\begin{figure}[t]
\vskip 10.2truecm
\includegraphics{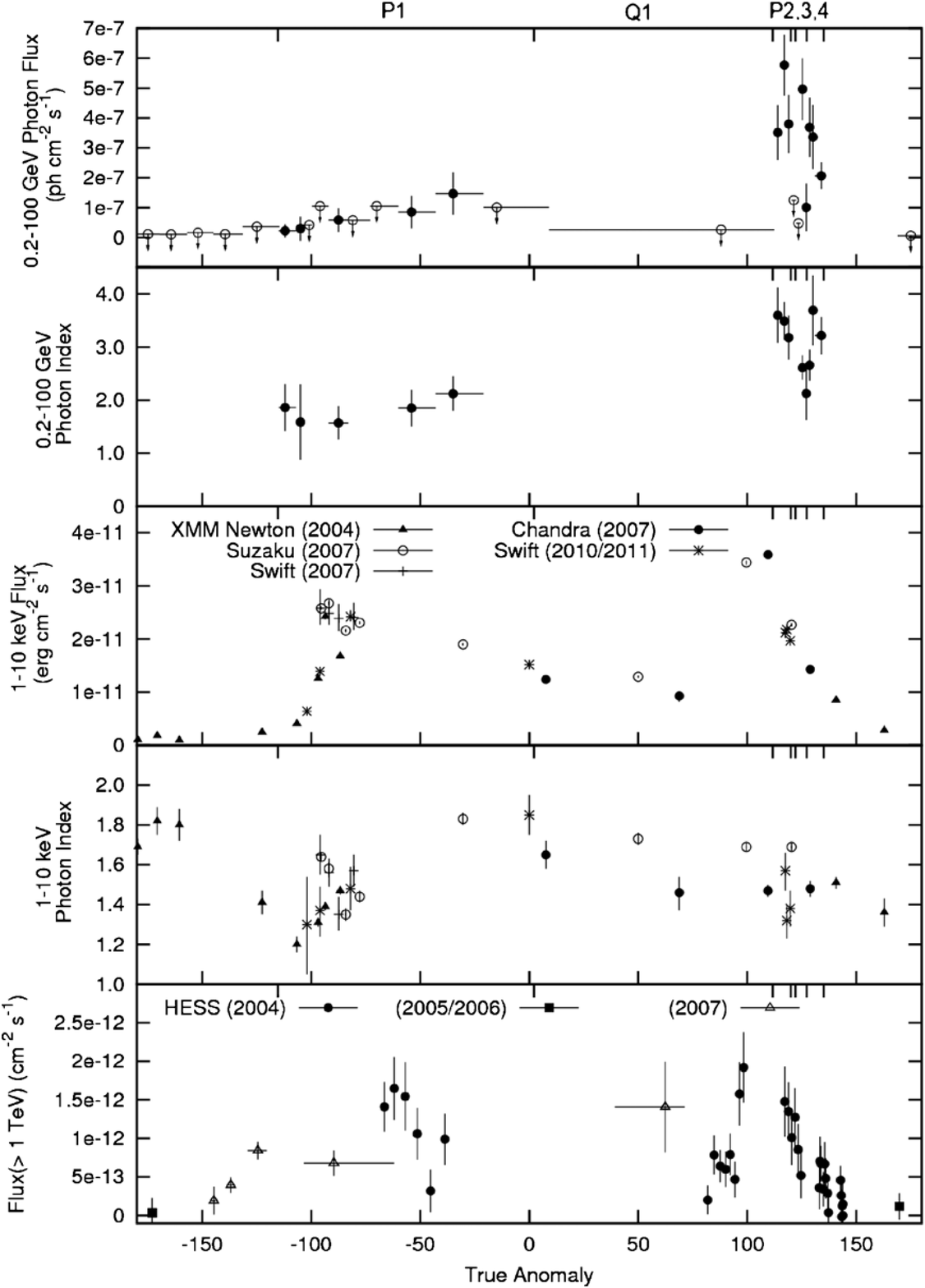}
\caption{The observations of the binary system PSR1259-63/SS2883
in the X-rays, GeV and TeV $\gamma$-rays close to the periastron passage (from~\cite{tam11b}, reproduced by permission of the AAS). Note that the GeV emission (with the maximum at about a month after periastron) seems not to be correlated with the X-ray or the TeV emission.}
\label{fig7}
\end{figure}

An observational breakthrough has come with the next generation of the Cherenkov telescopes, which were able to discover clear TeV $\gamma$-ray signals from three massive binary systems, LS2883/PSR1259-63~\cite{ah05b}, LS5039~\cite{ah05c}, and LSI 61 303~\cite{al06}.
The signal in the first system was detected close to the periastron passage of the compact object which orbit around the massive star has well known parameters~\cite{ah05b,abdo11c}. 
The observations of LS2883/PSR1259-63 with the Fermi-LAT telescope show quite unexpected features such as the brightening of GeV emission about a month after the periastron passage~\cite{tam11b}. This GeV emission seems not to be directly correlated with that in TeV since its maximum appears at latter period (see Fig.~7 for a collection of the observations close to the periastron of X-rays, GeV and TeV $\gamma$-rays).
LS5039 and LSI 61 303 showed clear modulation of the TeV signal with the periods of the binaries. The maximum emission corresponded to the location of the compact objects in front of the massive stars~\cite{al09,ah06e}. The $\gamma$-ray spectrum of LSI 61 303 is well described by a single power law without evidence of a spectral cut-off~\cite{al06}. On the other hand, the spectrum of LS5039 is described by a single power law at the low state but by the power law with an exponential cut-off at the high state~\cite{ah06e}. 
A clear correlation of the TeV and X-ray emission from the TeV $\gamma$-ray binaries have been reported, thus supporting their production by the same population of electrons, e.g. LSI 61 303~\cite{an09b}, LS 5039~\cite{tak09}, and also HESS J0632+057~\cite{bon11}. 
These binaries have been also detected in the GeV $\gamma$-rays showing clear anticorrelation with the TeV light curve~\cite{ab09e,ab09f}. It is know at present that the TeV emission, at least from LSI 61 303, shows a long time scale variability during which the light curve changes significantly its shape (see Fig.~8 on the right~\cite{acc11,jog11}). A comparison of the spectra measured from LS5039 and LSI 61 303 in the GeV and TeV energy ranges shows two distinct components, the first extending above and the second below $\sim$10-100 GeV (see Fig.~8 on the left~\cite{ab09e,ab09f}. Such a feature was not expected by any theoretical model. Recently, two additional massive binary systems, HESS J0632+057 detected in the TeV energies~\cite{ah07c,acc09} and 1FGL J1018.6-5856 detected in the GeV energies~\cite{ack12}, 
show modulation of the $\gamma$-ray signal with the periods of their binary systems. 

\begin{figure*}[t]
\vskip 6.5truecm
\includegraphics{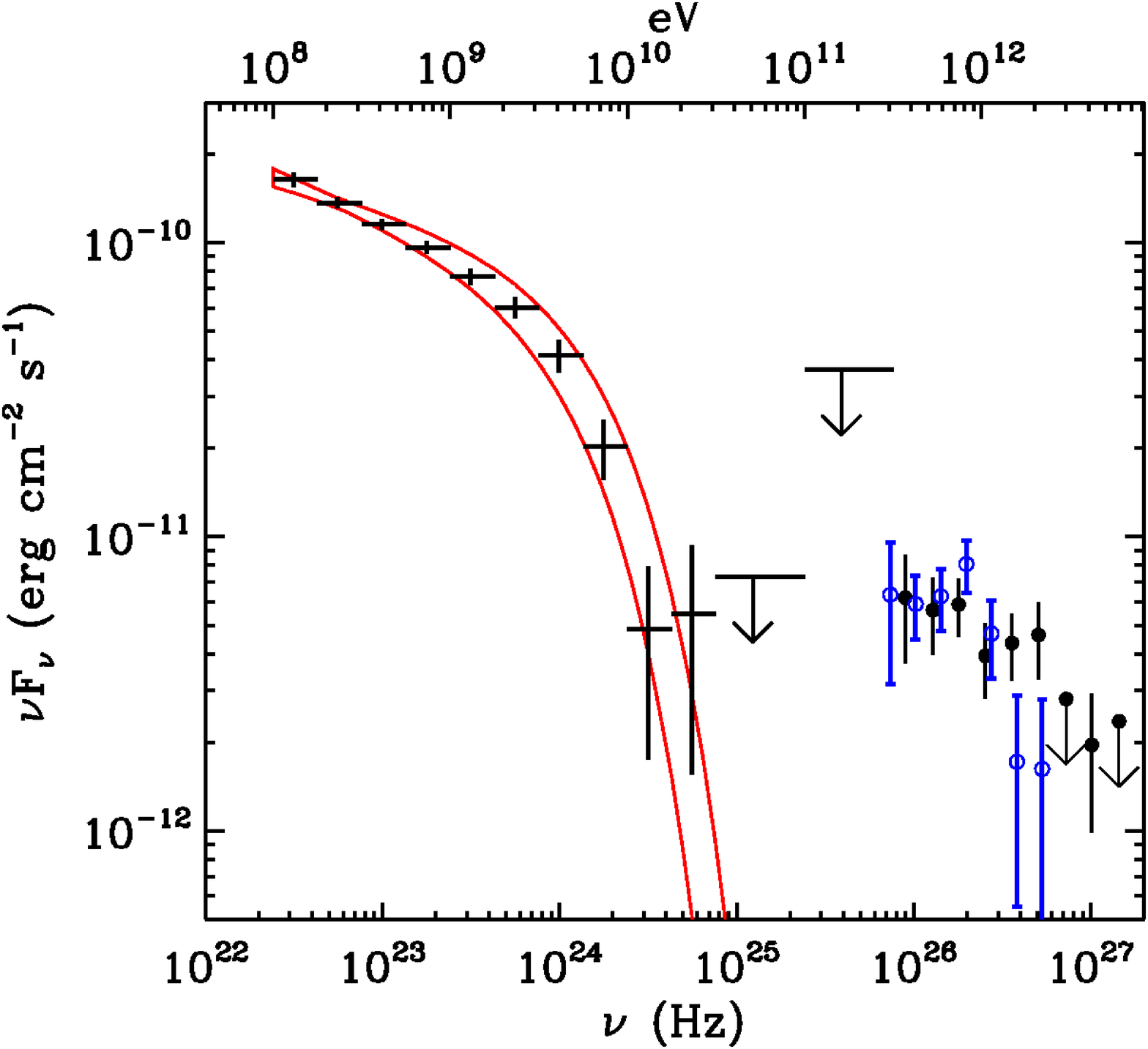}
\includegraphics{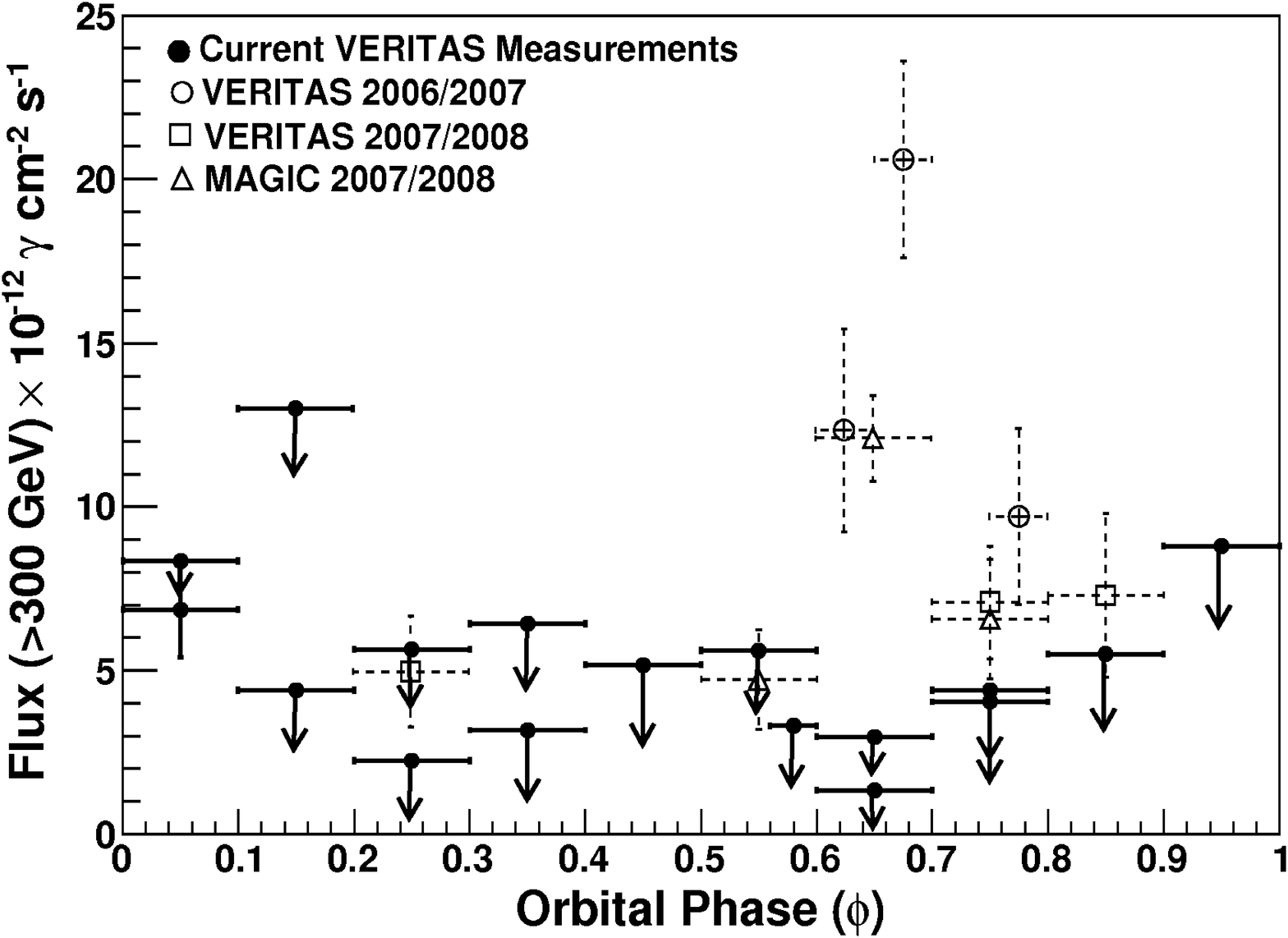}
\caption{On the left: Energy spectrum of LS I 61 303 (Abdo et al.~2009f, reproduced by permission of the AAS). Note the clear deficiency of the $\gamma$-ray emission at energies between $\sim$10-100 GeV. On the right: Light curves of LS I 61 303 reported by different Cherenkov telescopes (from~\cite{acc11}, reproduced by permission of the AAS). There is no detection of the TeV emission at phases $\sim$0.6-0.8 by VERITAS in 2008-2010.}
\label{fig8}
\end{figure*}
\subsection{Present understanding of the $\gamma$-ray emission from 
binary systems}

The X-ray emission observed from the TeV binaries is atypically weak with respect to their $\gamma$-ray emission, 
in contrast to the observations of other accreting X-ray binaries. Moreover, their radio morphologies suggest that the compact objects in these binaries are rotation powered pulsars~\cite{dub06a}. Therefore, the most popular scenario, which tries to explain the emission features from these objects, assumes an existence of a 
young energetic pulsar as already observed in the case of the binary system LS2883/PSR1259-63.
The interaction of the pulsar wind with the strong wind from the massive star is responsible in this scenario for the formation of the shock wave within the binary system. Particles can be accelerated at this shock to 
at least TeV energies. Alternatively, they can be injected from the inner pulsar magnetosphere or accelerated in the pulsar wind region as expected in the case of isolated pulsars (see Sect.~2 and 3). 

Recently, a magnetar-like X-ray burst was observed from the direction of LS I 61 303, suggesting that the pulsar can be characterised by a very strong surface magnetic field~\cite{th12}. In fact, it was shown that particles could be accelerated to TeV energies also in such accreting magnetar scenario~\cite{bed09b,th12}.

\subsubsection{Leptonic models}

In the most popular scenario, $\gamma$-rays in TeV binaries are produced by relativistic electrons which comptonize the thermal radiation from the massive companion (e.g.~\cite{mt81}). In fact, the optical depths within the massive binaries are clearly above unity for electrons with energies above a few tens of GeV. Thus, high energy $\gamma$-rays are expected to be produced with large efficiency (e.g.~\cite{ps87,mos93}). However, in such a case their absorption becomes also important. 
The optical depths were calculated for the massive stars in LS 5039 and LSI 61 303 by e.g.~\cite{bed06a,db06,dub06b}.
Due to large values of the $\gamma$-ray optical depths, it is expected that the $\gamma$-ray spectra from these binaries are formed in the IC $e^\pm$ pair cascades propagating in the anisotropic radiation field of the massive star as already considered some time ago by Bednarek~\cite{bed97}. The calculations of the spectra produced in the IC $e^\pm$ pair cascade model, which escape from these binaries, have been performed in several papers, see e.g.~\cite{bed06b,bed07,ah06f,st07}. These models predict clear anticorrelation of the GeV and TeV emission (e.g.~\cite{bed06a}). This general feature has been later confirmed by the GeV-TeV observations of LS5039 and LSI 61 303. However, the basic conditions for the cascade process within the binary system are not well known.
For example, the energy losses of the primary electrons and the secondary $e^\pm$ pairs in the cascade on the synchrotron radiation can play an important role (see~\cite{bed97,bg07,bos08,kha08}. It was shown in these studies that, if the magnetic field in the cascade region is above $\sim 1$ G, a significant energy is transferred from leptons to synchrotron radiation. As a result,
IC $e^\pm$ pair cascade starts to be inefficient especially at TeV energies when the IC cross section in the Klein-Nishina regime drops significantly. 
By comparing the energy loss rates of electrons on the synchrotron and the IC processes, we can estimate the limit on the stellar magnetic field strength below which the IC losses start to dominate.
When the ICS occurs in the Thomson regime, it is simply given by the condition $B < 40T_4^2$ G, where $T = 10^4T_4$ K is the surface temperature of the companion star~\cite{bed97}. 
For typical surface temperatures of stars in these binaries, of the order of $3\times 10^4$ K, this critical value of the magnetic field is equal to $\sim 400$ G.
Note however, that the energy density of magnetic field around the star drops in the interesting range of distances as $U_{\rm B}\propto R^{-4}$, whereas the energy density of stellar radiation drops only according to $U_{\rm B}\propto R^{-2}$.
Therefore, we conclude that the region, where the synchrotron losses of electrons dominate over their IC losses, should be confined to that relatively close to the companion star. Provided that the cascade process does not occur deep in the Klein-Nishina regime, the synchrotron energy losses do not affect it strongly  when the compact object is close to the apastron (the phase of the maximum of TeV emission), i.e. at the distance of $4-15$ stellar radii. At these distances from the companion star, the magnetic fields is likely to be of the  order of $\sim$1 G, in accordance with the calculations in~\cite{bed97,bos08,kha08}.

The magnetic field of a massive star can also change the directions of the primary and the secondary leptons resulting in different cascade scenarios. The following cases can be distinguished: (1) linear cascade in which particles ($\gamma$-rays and leptons) follow the direction of the primary electron (e.g.~\cite{cer09}; (2) completely locally isotropized cascade in which the secondary leptons are isotropized by the local magnetic field in their place of birth~\cite{bed06b,bed07,ah06f}; and the most general case (3) magnetically driven cascade in which the secondary leptons are driven by the local magnetic field with a certain structure~\cite{sb05}. 
Each of the above case predicts different angular distribution of the escaping $\gamma$-ray photons on the sky, thus resulting in different light curves. Note that all these cascade scenarios can in principle be realized in different parts of the same binary system. 

Another important factor affecting on the escaping $\gamma$-rays may be a variable, asymmetric, and clumpy wind from the massive star.
A change of the stellar wind pressure should result in the change of the location of the shock within the binary system. Thus, the conditions for the acceleration of electrons and the development of the IC $e^\pm$ pair cascade may change significantly, influencing the $\gamma$-ray light curve. An aspherical wind, with its slow and dense equatorial part but fast and low density polar part, may result in a drastic change of the location of the shock with respect to the companion star at different phases of the compact object.
This may strongly affect the conditions for the scattering of the soft radiation by electrons and the development of the IC $e^\pm$ pair cascade~\cite{sb08,st09,bog12}. Note that also the pulsar wind is expected to be aspherical (e.g.~\cite{bog99}). Therefore, the geometry of the $\gamma$-ray production in the IC $e^\pm$ pair cascades initiated by electrons in the vicinity of the massive star may be very complicated.

The winds of the massive stars in TeV binaries are expected to be clumpy. Such dense clumps provide the medium for the propagation of relativistic particles with very different properties (e.g. magnetic field strength, density of matter). This might result in a more complicated shape of the observed $\gamma$-ray spectrum~[\cite{zdzir10}. Moreover, the $\gamma$-ray light curve
of the binary system can be also influenced by the relativistic flow of the plasma along the cometary tail of the pulsar on its orbit around the companion star~\cite{dub10}. This effect is not expected to be very strong due to a rather slow pulsar wind behind the shock (velocities of the order of $0.3-0.5$ of the velocity of light).
The complicated shapes of the GeV-TeV spectra observed from LS 5039 and LSI 61 303 suggest the existence of two populations of electrons or two different emission mechanisms. In fact, an essential part of the GeV component could come from the inner pulsar magnetosphere. On the contrary, the TeV component might be produced by electrons accelerated in the pulsar wind region~\cite{to11}. In another scenario,
the two populations of relativistic electrons can be produced within the binary system as a result of the acceleration of particles on the two shock structure separated by a contact discontinuity. Such two shocks appear in the winds from of the pulsar and stellar side~\cite{bed11}. The conditions at both shocks can differ significantly (magnetic field strength, acceleration efficiency) resulting in different acceleration efficiencies of particles and their energy loss rates. It
has been shown that electrons accelerated on both shocks can reach different maximum energies saturating at $\sim$10 GeV at the shock from the side of the massive star and up to several TeV at the shock from the pulsar side~\cite{bed11}.
In summary, it is not surprising that precise predictions of the behaviour of the $\gamma$-ray light curves from the massive binaries are very difficult and they are not well described by the present models. The light curves and spectra may change in time due to not well known conditions within the binary systems discussed above. They depend on many parameters which at present are not constrained with enough precision.

\subsubsection{Hadronic models}

The production of $\gamma$-rays in the massive binaries in the  hadronic processes seems to be less likely due to a relatively long energy loss time scale of relativistic hadrons in the hadronic collisions with the matter of the stellar wind (for the expected wind mass loss rate of the order of a few $10^{-7}$ M$_\odot$ yr$^{-1}$) with respect to their escape time scale with the plasma flowing from the binary system along the shock structure. However, such models have also been considered (e.g.~\cite{rom03,rom05,kaw04,cher06}). 
Definitive answer to the question of the importance of hadronic $\gamma$-ray production in TeV binaries will come from observations with large scale neutrino telescopes by which the predicted neutrino rates might be detectable (e.g.~\cite{ch06,th07,nr09}).

\subsection{Binaries with pulsars - prime targets for CTA}

Unexpected features of the $\gamma$-ray emission from the TeV binaries have been already well documented by the present Cherenkov and satellite observations. The next generation of the Cherenkov arrays should be after all able to establish whether the presently observed TeV binary systems are persistent
sources of the modulated $\gamma$-ray signal. If also TeV binaries other than LS I 61 303 show long time variability, what features determine the high level of the $\gamma$-ray emission? Are these effects related in some way to the conditions created by the companion stars, e.g. similar to the long term activity observed in the Sun (e.g. the change of the stellar wind parameters - density, velocity, magnetic field - stellar cycles)?
This problem could be answered by the multiwavelength observations in the high energy domain together with the observations in the radio to optical range. A precise determination of the $\gamma$-ray light curves and the measurement of the $\gamma$-ray spectra at least in the Cherenkov energy range (sub- to multi-TeV) with an order of magnitude better  sensitivity of the CTA will allow them for a more detailed testing of different cascade scenarios for the $\gamma$-ray production mentioned in Sect.~5.2. 

CTA should also be able to measure shapes of the $\gamma$-ray spectra at a few tens GeVs, i.e. in the region of the dip separating the high energy and very high energy components. Such detailed spectra at the sub-TeV energies should enable theoretical studies of the origin of this feature (is it due to the presence of different components or other phenomena such as e.g., the absorption effects of $\gamma$-rays in the UV-X-ray radiation ?). 
CTA will be able to make precise measurements of the phase dependent spectra from a much larger sample of the TeV binaries. Thus, it will allow us to investigate of the acceleration 
processes of particles at different conditions at the shock region. 

The TeV spectra, reported from the $\gamma$-ray binaries, 
extend up to at least a few TeV without any evidence of the final cut-off. A detection of such cut-offs would allow us to constrain maximum energies of the accelerated particles. If these maximum energies extend up to a hundred TeV then the production of $\gamma$-rays by hadrons is preferable since electrons in the environment of the binary system (strong magnetic and radiation fields) would have problems with overcoming the radiation energy losses. Therefore, the measurements of the high energy cut-offs in the $\gamma$-ray spectra of these binaries with CTA would be of great importance for constraining the model parameters.

The models of $\gamma$-ray production in the pulsar winds
predict the existence of a very hard $\gamma$-ray component (spikes) being as a result of the comptonization of the external radiation field by leptons in the relativistic pulsar wind (e.g.~\cite{ba00,bk00,kha11}.
More recently, Cerutti et al.~\cite{cer08} compared the $\gamma$-ray emission expected in the case of LSI 61 +303 and LS 5039 to the TeV observations of these two sources. It was concluded that the energies of $e^\pm$ pairs in the pulsar winds cannot exceed 10-100 GeV. Future searches of such features in the spectra of TeV binaries (maybe only in some specific phases of the binary system?) with CTA (much better energy resolution and sensitivity in broader energy range) should enable unique determination of the basic parameters of the pulsar wind such as, e.g. its Lorentz factor, or the number of relativistic $e^\pm$ pairs. 

Sensitive observations of the binary systems at sub-TeV energies with the next generation CTA might also make possible a detection of the pulsed component originating in the inner magnetospheres of the binary pulsars as observed in the case of the Crab pulsar.  
In the extremely compact binaries, containing a pulsar and the WR type star, this pulsed $\gamma$-ray emission could be additionally modulated with the period of the binary system due to geometrical effects in the IC scattering process.
Therefore, searches for new TeV binaries, containing energetic pulsars and WR type stars, such as that present in Cyg X-3, will be of great importance. A comparison of the emission features of different objects  will enable a determination of the crucial parameters determining the $\gamma$-ray emission from these sources. 

Stellar and pulsar winds from $\gamma$-ray binaries should produce large scale nebulae around the massive binary systems, filled with energetic particles accumulated during the lifetime of the pulsar. In addition to the observed $\gamma$-ray emission modulated with the period of the binary, such nebulae could also produce a steady low level $\gamma$-ray emission~\cite{bb11}. Detection of this low level persistent TeV emission with CTA will be of great importance allowing us to constrain the history of the activity of this type of sources.

It is difficult to reach clear conclusions on the number of $\gamma$-ray emitting binaries within the Galaxy 
based only on the evolutionary studies due to unknown role of the energy output form the pulsar on the evolution of the companion star. Let us hope that the CTA will discover much more such objects during its scans of the galactic plane enabling population studies of the TeV binary systems in the future.

In the case of extended binaries, of the LS2883/PSR1259-63 type, a detection of the persistent and pulsed $\gamma$-ray emission at phases farther from the periastron passage would allow us to investigate the direct role of the pulsar in the emission process when it is also close to the companion star. Therefore, CTA should try to measure sub-TeV emission from PSR1259-63 when it is far away from the massive star. 
It will be interesting to find out whether the flaring GeV component, discovered recently by Fermi-LAT about a month after the periastron passage, has persistent nature and similar features at different passages of the pulsar. A discovery of an accompanying sub-TeV component by CTA at these phases would make possible to constrain the particle spectra responsible for this component and put some limits on the emission region within the binary system (e.g. constrain the possible shock extension to regions very close to the companion star surface as predicted at some phases of the binary by Sierpowska-Bartosik \& Bednarek~\cite{sb08}.   
This would provide new information about the nature of this surprising emission - whether it is due to an instantaneous state of the medium around the companion star or rather it is related to the properties of the pulsar and its wind, e.g. due to the geometrical effects in the collimation of the wind at specific directions?

\section{Microquasars}

The matter accreting onto a compact object such as neutron star or a
black hole is expected to form an accretion disk. In the case of the accreting NS a part of the matter from the disk can be expelled forming a jet which can move with relativistic velocities.  A similar jet is expected when the accretion occurs onto a
rotating Black Hole (BH). Then, the rotating BH is expected to provide efficient mechanisms for energy extraction~\cite{bz77,lov76,ble76}.
Jets are also expected to be accelerated and collimated hydromagnetically in the inner parts of the accretion disks~\cite{liv99}. An important consequence of these processes is the production of jets
which moving relativistically 
in the direction perpendicular to the plane of the accretion disk.
In fact, the large scale radio jets, some of them also showing features of the superluminal motion, were discovered in the galactic X-ray binaries~\cite{mir92,mir94,tin95,hr95}. Up to now, several such objects have been classified as  microquasars in analogy to the quasar phenomenon observed around supermassive BHs in the nuclei of massive galaxies. Due to this analogy and also the evidence of the non-thermal radio to X-ray emission, indicating the presence of TeV electrons (see e.g. observation of X-ray synchrotron radiation at parsec distances~\cite{cor02,cor05}, microquasars have been suspected to be sites of acceleration of particles able to produce $\gamma$-rays in the GeV-TeV energy range. 

\subsection{Observations of Gamma-rays from microquasars}

\begin{figure}[t]
\vskip 7.truecm
\includegraphics{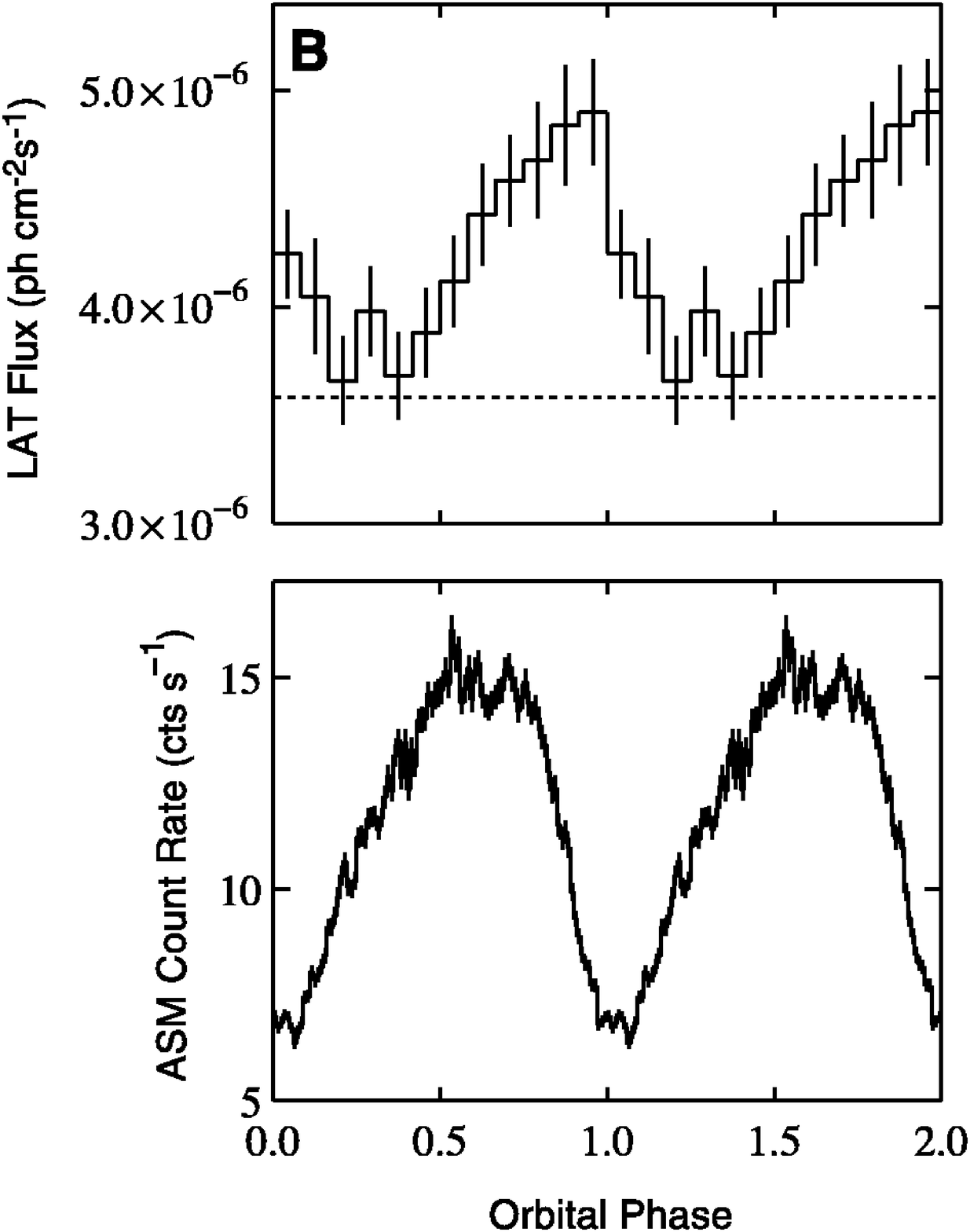}
\caption{Fermi $\gamma$-ray light curve (during the outburst) and the RXTE ASM X-ray light curve (entire lifetime of RXTE) observed from Cyg X-3 binary system~\cite{ab09g}, reproduced by permission of the AAAS). Note the anticorrelation between the GeV and X-ray emission.}
\label{fig9}
\end{figure}

The possible detection of the $\gamma$-ray emission from the well known binary system Cyg X-3, belonging to the microquasar class, were already claimed in data from the past satellite telescopes~\cite{lam77,mor97}, but see~\cite{her87}. 
Only recently, the GeV emission was reported in~\cite{tav09,ab09g}. The observed $\gamma$-ray emission is clearly transient~\cite{tav09,ab09g,bul11}. It appears close to the quenched radio state just before the major radio flares observed in this system. The spectrum above 100 MeV is well described by a single power law with the differential spectral index $-2.7\pm 0.05(stat)\pm 0.20(syst)$ and the peak flux $\sim 2\times 10^{-6}$ photon cm$^{-2}$ s$^{-1}$  in the case of Fermi-LAT telescope~\cite{ab09g} and by the spectral index $-2.02\pm 0.28$ in the case of AGILE telescope~\cite{bul11}. 
The emission shows clear modulation with the orbital period of the binary system, with the maximum close to the location of the compact object behind the companion star~\cite{ab09g}. This is also the case of the TeV  binaries.
The $\gamma$-ray light curve is clearly anticorrelated with the X-ray light curve (see for details Fig.~9)  
Due to these similarities, it might be supposed that Cyg X-3 should also emit TeV $\gamma$-rays at phases when the compact object is in front of the companion star. Up to now, the TeV $\gamma$-ray emission has not been discovered from Cyg X-3, in spite of the extensive observations with the MAGIC telescope~\cite{ale10}. However these observations start to constrain the $\gamma$-ray spectra observed at low energies (see Fig.~10).

\begin{figure}[t]
\vskip 7.5truecm
\includegraphics{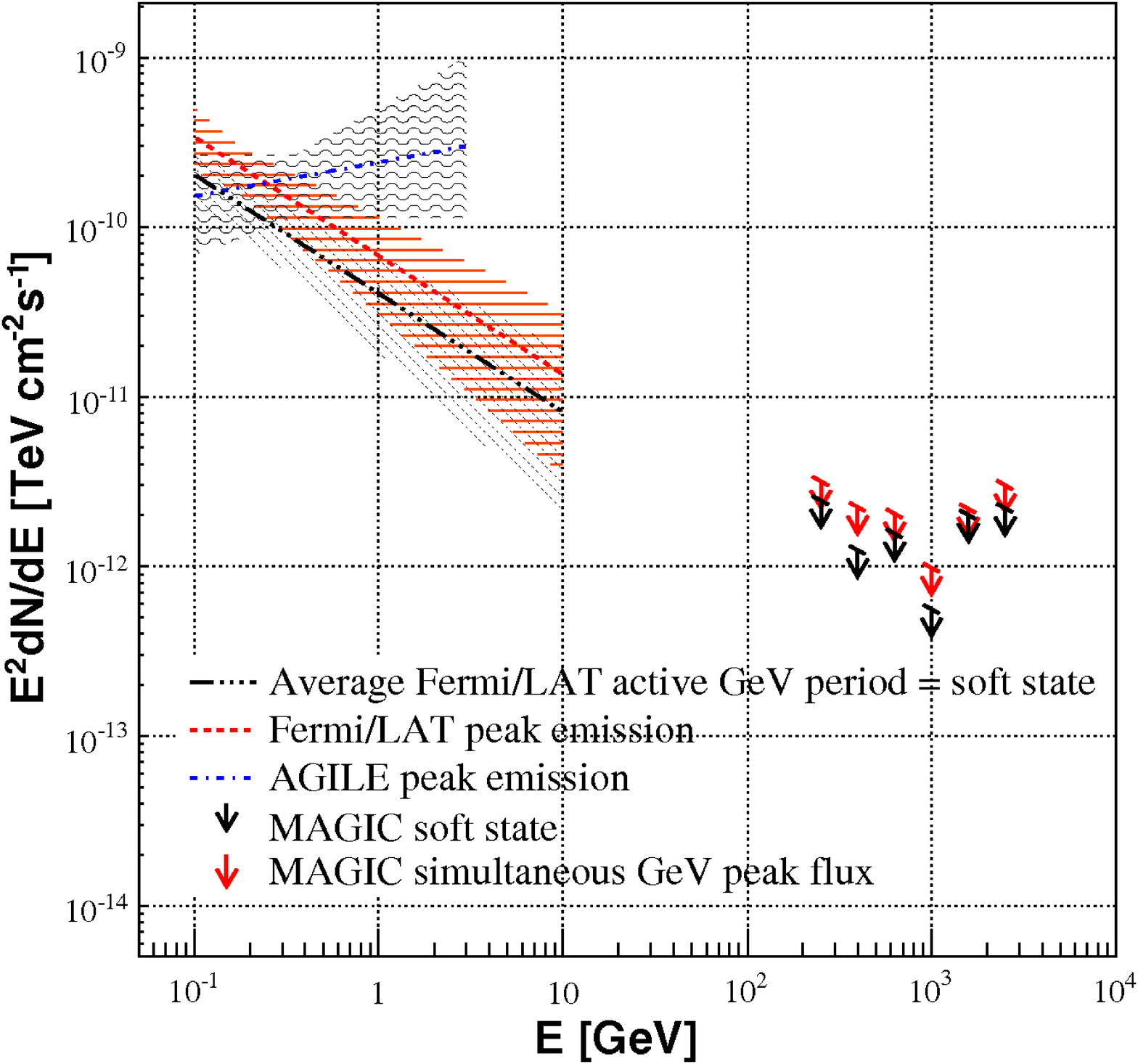}
\caption{The $\gamma$-ray spectrum from Cyg X-3 measured by Fermi (Abdo et al.~2009g), AGILE (from~\cite{bul11} and the upper limits from MAGIC~\cite{ale10}, reproduced by permission of the AAS). Note that the 
upper limits from MAGIC start to constrain Cyg X-3 spectrum at TeV energies.}
\label{fig10}
\end{figure}

In the case of another microquasar, within the massive binary system Cyg X-1, there is also evidence of a transient GeV-TeV emission~\cite{al07,sab10}. A flare of the TeV emission from Cyg X-1 occurred close to one of the flares in the hard X-rays observed within the 3 day flaring episode (e.g.~\cite{mal08}). It lasted for less than an hour when the compact object was behind the massive companion. The significance of this flare is rather low (below $5\sigma$). Also a separate flare above $100$ MeV, on a time scale of 1-2 days, was reported recently by the AGILE Team~\cite{sab10}. The GeV flares seem to be rather exceptional during the 300 day observation period with the AGILE telescope~\cite{dm10}. Unfortunately, the GeV flare has not been detected by Fermi-LAT telescope during this same period.
The $\gamma$-ray emission from microquasars in the low mass binary systems have not been discovered up to now in spite of extensive observations, see e.g. GRS 1915+105~\cite{ac09,sai09} or SS433~\cite{ah05d,hay09,sai09}.

\subsection{Theoretical modelling and expectations}

It is generally expected that the high energy $\gamma$-ray emission in the
previously discussed two microquasars is related to the presence of a massive star close to a mildly relativistic jet. 
A few different sites within the massive binary systems have been discussed as responsible for the origin of $\gamma$-rays. 
Close to the base of the jet, where the magnetic field strength is expected to be large, nuclei are preferentially accelerated to energies enabling the GeV-TeV $\gamma$-ray production. However, electron acceleration is expected to be saturated there at relatively low values (GeV range) due to the synchrotron energy losses, thus preventing the TeV $\gamma$-ray production. On the other hand, the acceleration of nuclei should not be strongly influenced by their energy losses allowing them to reach even PeV energies. 
Nuclei with such energies may interact efficiently with the X-ray photons of the inner parts of the accretion disk producing $\gamma$-rays and neutrinos (e.g.~\cite{lw01}). At intermediate distances from the accretion disk, its radiation field is still strong enough for an efficient disintegration of nuclei but it is too weak for the pion production in collisions with photons~\cite{bed05b}. In this case, neutrons released from the nuclei, can outshine the accretion disk producing $\gamma$-rays and neutrinos in collisions with the matter of the disk (see e.g.~\cite{nel93}).
At larger distances from the disk, the radiation field of the companion star starts to dominate over the radiation field of the accretion disk. Then, neutrons from the disintegration of nuclei outshine mainly the massive companion producing neutrinos and $\gamma$-rays in collisions with the stellar atmosphere~\cite{bed05b}. At these distances the synchrotron energy losses of accelerated electrons become low enough, allowing them to be accelerated to TeV energies. Therefore, electrons can efficiently produce $\gamma$-rays by the inverse Compton scattering of the stellar radiation (e.g.~\cite{geo02,rom02},
and recently~\cite{zdzir11}.
The outer regions of the jet (which are still within the binary system) seem to be likely sites for the production of GeV-TeV emission which fulfills the observational characteristics obtained from the observation of the Cyg X-3 binary system, i.e. the modulation of the GeV signal with the period of the binary system. However, $\gamma$-ray production in the interaction of particles with the radiation and matter of the accretion disk can also be 
an important mechanism, provided that the observed low level of modulation of the $\gamma$-ray signal is due to other effects, e.g. the precession of the disk and/or the jet with the period of the binary system.

The high energy $\gamma$-rays produced in the compact binary systems, containing massive stars (such as Cyg X-1 and Cyg X-3), should also suffer from the strong absorption in the radiation field of the star (e.g.~\cite{mos93}). As a result, the spectra of the escaping $\gamma$-rays are formed in the IC $e^\pm$ pair cascade, developing in the anisotropic radiation field of the massive companion (as in the case of TeV binaries, see Sect.~5.2.1). The detailed calculations of the optical depths of the massive star environments in the binary systems Cyg X-1 and Cyg X-3, show that the escape of TeV $\gamma$-rays depends strongly on the phase of the binary system, see e.g.~\cite{bg07} for Cyg X-1 and~\cite{bed10} for Cyg X-3. It was shown that the strong TeV $\gamma$-ray emission from Cyg X-1 at the phase reported by the MAGIC observations i.e.
at the location when the compact object is behind the massive star~\cite{al07}, is rather unlikely~\cite{bg07}. The injection spectrum of electrons in this case should be very hard, as shown in~\cite{zdzir09}. Due to the very strong $\gamma$-ray absorption close to the hot WR type stars, it is also rather unlikely to expect the present Cherenkov Telescopes to detect any emission from Cyg X-3~\cite{bed10}.
In principle, the $\gamma$-ray emission observed
from Cyg X-3 might originate also as a result of the ICS of the disk radiation by relativistic electrons in the inner jet~\cite{cer11}. Detailed calculations of the $\gamma$-ray spectra along such general scenario, which include the development of  the anisotropic IC $e^\pm$ pair cascade in the disk radiation, has been recently made by Sitarek \& Bednarek~\cite{sb11}. It is concluded that the observed $\gamma$-ray spectrum from Cyg X-3 can be explained provided that the emission region in the jet is at the distance up to $100-300$ inner radii from the accretion disk. In this case, the observer should be located at a relatively small angle to the jet axis.

Relativistic particles may appear in the binary system due to acceleration on the shock waves being a result of the interaction of the jet with a dense, clumpy wind from the massive star (e.g.~\cite{pbr08,rom10,dub10b}. In such scenario $\gamma$-rays might be produced in hadronic collisions, as considered by e.g.~\cite{ar09,ow09}. The $\gamma$-ray fluxes strongly depend on the properties of the stellar wind (density, velocity) which are at present not well known. Also the interaction of the jet in its termination region (on a parsec scale) may result in the production of a steady, high energy $\gamma$-ray emission with an intensity difficult to determine reliably, e.g.~\cite{aa99,bor09}. However,
a discovery of such a persistent GeV-TeV $\gamma$-ray emission would provide important constraints on the content and energetics of the microquasar jets. 

Originally, the TeV $\gamma$-ray binaries (see Chapter.~5) have been considered to belong to the microquasar class and their early modelling postulated the injection of relativistic particles along the jet (e.g.~\cite{db06,bed07}.
However, now the hypothesis of the presence of energetic pulsars in these systems is more widely accepted.

\subsection{Open questions to be answered by CTA}

Interesting questions concerning the acceleration of particles, their propagation and radiation process in microquasars may be put in new light with an order of magnitude more sensitive observations above a few tens of GeV.  
Long term monitoring at sub-TeV energies should confirm the observed relation of the $\gamma$-ray emission only to the major radio flares of Cyg X-3. However, the question also appears whether lower level $\gamma$-ray emission is produced also in the quiescent radio stages? 
In the case of the high mass microquasars, such as Cyg X-3
and Cyg X-1, the localization of the emission region within the binary system is of crucial importance. An observation of a break in the $\gamma$-ray spectrum at a few tens of GeV would be consistent with the absorption by the stellar radiation in the system. However, the predicted fluxes on the level of $\sim 1\%$ CU at 20-30 GeV and $0.1\%$ CU above 100 GeV~\cite{bed10} would be probably difficult to detect even with CTA particularly that the emission has a transient nature with the duty time of the order of a few days. A cut-off in the spectrum at even lower energies might suggest $\gamma$-ray absorption by the radiation of the hot inner accretion disk, indicating the localization of the emission region in the inner jet. It will be interesting to search with CTA for an anticorrelation of GeV and (possible) sub-TeV fluxes as observed in the case of TeV binaries. Their discovery would show an importance of the $\gamma$-ray absorption and a formation of "escaping" $\gamma$-ray spectra in the cascading processes.
A confirmation of the detection of the TeV emission from Cyg X-1 would be of great importance since stars in Cyg X-1 binary system have much larger separation than those in Cyg X-3. Thus, the escape conditions for the TeV $\gamma$-rays produced in the binary system
are much less restrictive. 

CTA should be able to conclude whether microquasars in the low mass X-ray binaries (LMXBs) can also produce $\gamma$-rays.
The acceleration of particles to TeV energies in LMXBs is expected to occur also efficiently as concluded from the observations of synchrotron radiation in the X-ray energy range (e.g.~\cite{cor02,cor05}). This process is expected to be determined rather by the jet properties than by the companion star. The upper limits (or low level detections) of the GeV-TeV photons with CTA would allow us to answer the question on the importance of a massive star in the microquasar for high energy processes. For example, in the LMXBs $\gamma$-rays might be produced in another mechanism than that expected in the HMXBs, e.g. the Synchrotron Self-Compton process might dominate.

\section{Massive stars in the close binary systems}

Strong winds of massive stars (O and WR type) may collide producing a double shock structure separated by a contact discontinuity. These winds are characterised by large energies, large velocities and strong magnetic fields (due to the strong surface magnetic fields of the companion stars). It is expected that these shocks accelerate particles (electrons and hadrons) to energies allowing them to produce GeV-TeV $\gamma$-rays. These systems seem to be unique laboratories for the investigation of the mechanisms of particle acceleration (shock acceleration or reconnection of magnetic fields) due to the well defined  target for the relativistic particles within the binary system (density of matter, soft radiation field).

\subsection{High energy observations of massive binaries}

\begin{figure}
\vskip 6.truecm
\includegraphics{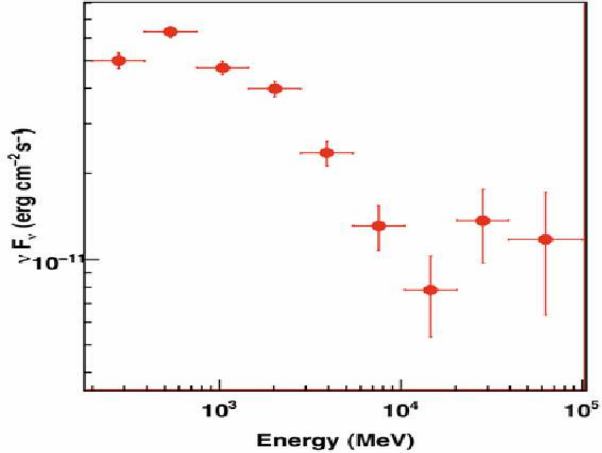}
\caption{The $\gamma$-ray spectrum observed from the direction of Eta Carinae binary system. Two spectral components are clearly observed,
the soft one, showing exponential cut-off at a few GeV and the hard one,
extending up to $\sim$100 GeV (from~\cite{wfl11b}, reproduced by permission of the A\&A).}
\label{fig11}
\end{figure}

Up to now only one massive binary system, Eta Carinae, has been found in the direction of the Fermi-LAT source at energies below $\sim$100 GeV~\cite{tav09,abdo10j,wfl11}. 
The high energy emission of this source is characterised by an intriguing two-component spectrum which can be approximated by two power laws, the first one with an exponential cut-off at a few GeV and the second extending up to at least $\sim$100 GeV (see Fig.~\ref{fig11}). 
Recently, it was shown that the high energy part of this emission must be associated with the Eta Carinae~\cite{wfl11b}. 
Whether this second component extends through the TeV energy range or cuts off earlier is at present unknown.  

It is also possible that a part of the TeV emission from the direction of the open cluster Westerlund 2 might originate in the massive binary system, WR 20a~\cite{ah07b}.
However, it is not yet clear whether any emission comes from the Westerlund 2 or from the nearby PWNa~\cite{abr11b}. Such PWNa is probably associated with the $\gamma$-ray pulsar recently detected by the Fermi-LAT~\cite{saz10}.
The upper limits on the TeV emission from other massive binary systems, containing WR type stars such as  WR 146 and WR 147, have been reported recently by the MAGIC Collaboration~\cite{al08b}. They are on the level below the theoretical predictions allowing one to constrain the parameter space of some models for their origin~\cite{rei06}.

\subsection{Models for $\gamma$-ray emission}

The high energy radiation, expected in terms of the wind collision model, was considered in more detail in several papers~\cite{eu93,br03,bed05a,rei06,pd06,wfl11b,bp11b}. In general,
the  acceleration process of electrons and hadrons can be saturated by the radiative  energy losses or by particle escape from the binary system. In the case of electrons, the synchrotron process and the Inverse Compton scattering on the stellar radiation play the dominant role. Electron acceleration can be saturated by the synchrotron losses at TeV energies.
However, electron acceleration can be also saturated by IC energy losses at much lower maximum energies, of the order of several GeV (for detailed formulae see e.g.~\cite{bp11b}). 
The above mentioned values are typical for binary systems containing two massive stars. Which of these saturation scenarios is realised in particular binary system, depends on the magnetic field and the radiation field of the massive star.
On the other hand, acceleration of hadrons can be saturated either by energy losses in hadronic collisions with the wind matter or due to the escape from the binary system along the shock.
The maximum energies of hadrons determined by hadronic interactions were estimated as a few hundred TeV (see, e.g.~\cite{bp11b}).
On the other hand, the maximum energies of hadrons, if their escape from the binary system is important, are expected to be of the order of a few PeV.
It is clear that depending on the conditions at the shock, the acceleration of electrons and/or  hadrons is saturated at very different maximum energies. This may lead to a variety of radiation scenarios depending on the phase of the binary system. They can be significantly different depending on the conditions in the binary systems. In order to put some light on the possible high energy processes occurring in the binary systems and on the acceleration mechanism of particles, only long term sensitive observations of these systems at high energies, simultaneous with those at other parts of the electromagnetic spectrum, are required. 

A few binary systems have been considered in more detail as possible $\gamma$-ray emitters (e.g. WR 140, WR 146, WR 147, WR 20a, Eta Carinae). In general, the models postulating the acceleration of electrons do not predict $\gamma$-ray emission  
at energies clearly above $\sim$100 GeV due to a saturation of the electron acceleration by radiative energy losses (e.g.~\cite{rei06,pd06}). On the other hand, hadronic models predict $\gamma$-ray emission which in principle can extend up to PeV energies~\cite{br03,bed05a,wfl11b,bp11b}. However, this emission, if produced in the binary system, should be strongly influenced by the absorption in the thermal radiation of the stars~\cite{bed05b,rei06,bp11b}. Moreover, $\gamma$-rays with TeV energies can be produced also by hadrons escaping from the binary system
into the relatively dense nebulae surrounding it (e.g.~\cite{ohd10b}). This emission should be steady in contrast to the $\gamma$-ray emission coming from the binary system which is expected to be modulated with the binary period due to the eccentric orbits of the stars and geometry dependent absorption of $\gamma$-rays with energies above $\sim$100 GeV. Therefore, various $\gamma$-ray emission features are expected from massive binary systems.  

\subsection{Future studies with CTA}

A variety of the emission features predicted by different scenarios can only be tested by more sensitive instruments such as CTA. CTA will be able to provide spectral information in the whole range of the binary phases i.e. at different conditions in the acceleration region. No detection of any $\gamma$-ray signal above $\sim$100 GeV would support their leptonic origin in the binary systems. However, a detection of $\gamma$-rays above $\sim$100 GeV, modulated with the binary period, would support the acceleration of hadrons within the binary. 

It would be interesting to determine whether the acceleration of electrons and/or hadrons occurs only at specific phases, depending on the conditions at the shock. 
Such investigations should throw new light on the conditions for acceleration of electrons or hadrons at shock waves.  
On the other hand, CTA could also discover a weaker steady $\gamma$-ray component from the surroundings of the binary systems containing two massive stars, thus enabling studies of the particle diffusion. This will put additional constraints on the acceleration process of particles at the shocks with different conditions.

\section{Accreting White Dwarfs}

Accreting magnetized White Dwarfs can provide a variety of conditions for acceleration of particles either in the inner, turbulent part of the accretion disk (the accretor or propeller phase) or in the thermonuclear explosions occurring on the surface of the WDs. Thermonuclear explosions may occur from time to time in the matter accumulated on the WD surface. The expelled matter creates a shock able to accelerate particles. 

\begin{figure}
\vskip 7.0truecm
\includegraphics{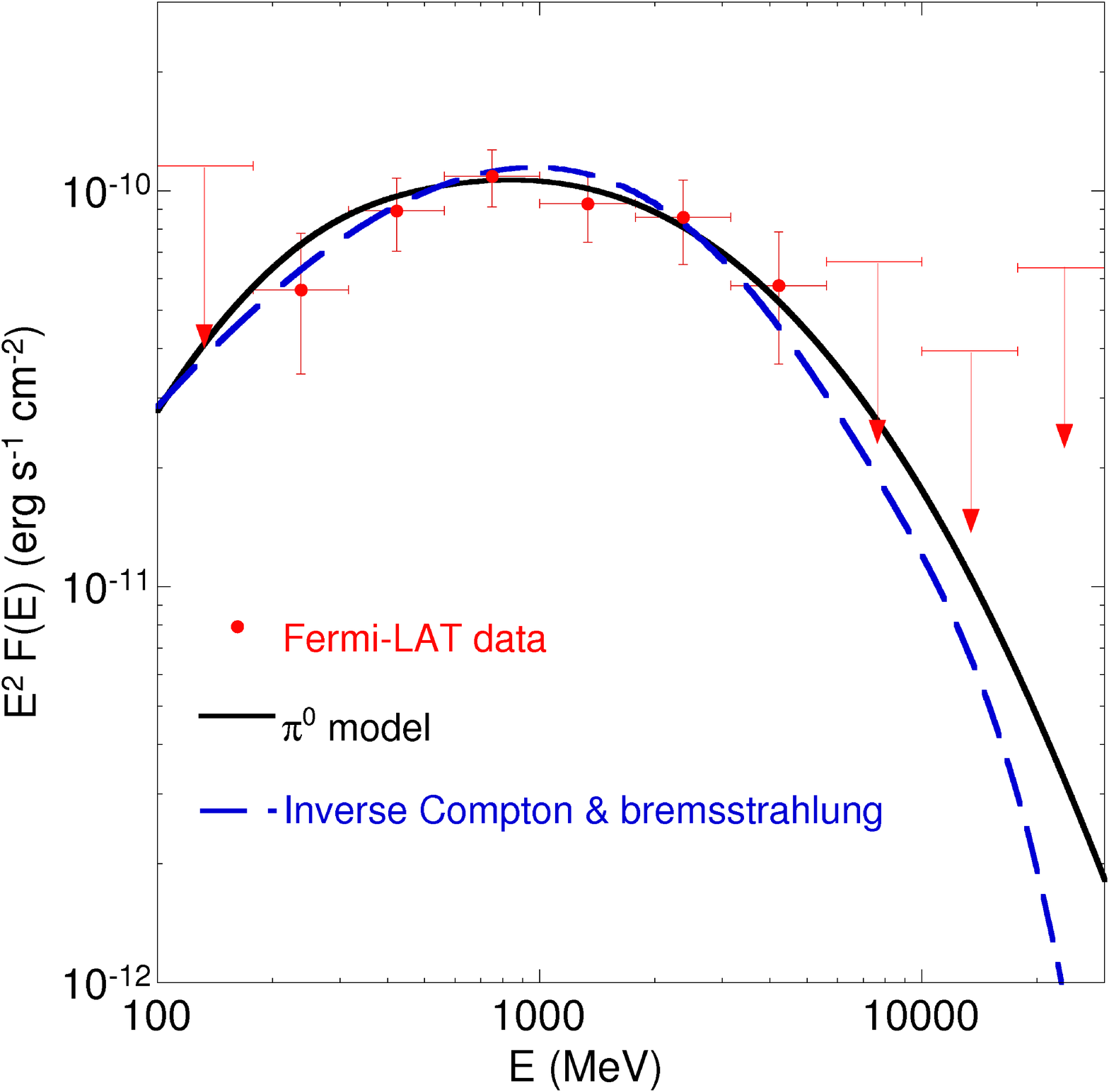}
\caption{$\gamma$-ray spectrum from the direction of the
symbiotic binary  V407 Cyg containing a White Dwarf on an orbit around the red giant (from~\cite{abdo10i}, reproduced by permission of the AAAS). Note that the spectrum seems to cut-off exponentially above $\sim$1 GeV.}
\label{fig12}
\end{figure}

Fermi-LAT has recently discovered $\gamma$-ray emission from the symbiotic binary, V407 Cyg,  containing a WD orbiting a red giant. The $\gamma$-ray flare is correlated with the optical Nova~\cite{abdo10i}.
The observed spectrum is hard below $\sim$1 GeV showing an exponential cut-off at larger energies (see Fig.~\ref{fig12}). An extrapolation of the spectrum (from the two last points with a simple power law) is above the 5 hr sensitivity of the planned CTA at energies above $\sim$100 GeV~\cite{ac11}. 

The mechanism of this $\gamma$-ray production (leptonic or hadronic) is at present unknown. 
The observed $\gamma$-rays are believed to be produced either by decay of pions produced by hadrons interacting with a dense wind of the red giant or by electrons which IC scatter the  soft photons from the red giant.
If it is leptonic, then additional high energy component in the spectrum could be present due to the accelerated hadrons. 
We conclude that future Nova explosions, although expected to be rather rare (the rate between $0.5-5$ yr$^{-1}$ in the Galaxy~\cite{lu11}, have a chance to be detected by CTA.
Such detections will put important constraints on the shock acceleration model at the conditions met in the symbiotic binaries and on the $\gamma$-ray radiation mechanisms. 

Recently, $\gamma$-ray emission from binary systems containing accreting WDs was studied in~\cite{mei00,bp11a}. In these models, it is proposed that the turbulent transition region between the rotating magnetosphere of the WD and the accreting matter can provide good conditions for the acceleration of particles (the propeller stage). These particles might be able to produce even TeV $\gamma$-rays in some systems, expected to be detectable by Cherenkov telescopes. However, accelerated electrons are probably not able to produce $\gamma$-rays with energies above $\sim$10 GeV due to the saturation of their acceleration by the synchrotron radiation.

\section{Other objects expected to emit TeV $\gamma$-rays}

The high energy processes in other Galactic objects 
have also been analysed in terms of the leptonic and hadronic models. These objects, e.g. isolated massive stars, White Dwarfs or Young Stellar Objects and accreting compact neutron stars (LMXBs) and White Dwarfs,
have been predicted to emit the $\gamma$-ray fluxes on the level detectable by the future Cherenkov telescopes.

\subsection{Winds of the classical massive stars}

Strong winds, produced by the early type stars (Be, O, WR), are expected to be non-stationary, containing smaller scale shocks. It has been argued that these shocks can accelerate particles either close to the stars~\cite{wc85,pol87} or at the terminal shocks of the stellar winds (e.g.~\cite{vf82,cm83}).
Electrons accelerated in such shocks have energies allowing them to produce of GeV $\gamma$-rays in the bremsstrahlung process~\cite{pol87}. On the other hand, hadrons accelerated to $\sim$100 GeV could produce $\gamma$-rays via decay of pions (e.g.~\cite{wc92}. The predicted fluxes seemed to be within the sensitivity of the EGRET telescope but no source of this type had been reported at that time. Massive stars usually stay close to their birth places which are within massive stellar associations. They are also expected to be sources of $\gamma$-rays due the presence of large scale shocks produced by a accumulation of stellar winds. 

In fact, the TeV emission was detected from the directions of a few open stellar clusters: Cyg OB2~\cite{ah02},
Westerlund 2~\cite{ah07b,abr11b}
and Westerlund 1~\cite{ohm10a}. Its origin is still debated. It is not excluded that a part of this emission comes from isolated massive stars. However, it will be rather difficult to identify the TeV emission from specific
massive stars within the open clusters even with the CTA angular resolution and sensitivity due to a large number of potential sources (several massive WR type stars, pulsars, supernova remnants, etc ...).
However, O and WR type stars, which are sometimes well outside clusters, can become interesting targets for the observations with CTA. Their possible detection will enable studies of the high energy processes occurring in the winds not far from stellar surfaces, constraining the structure and strength of the wind magnetic fields and high energy flaring activity in the magnetospheres of massive stars.

\subsection{Young stellar objects}

An other promising candidates are the
massive young stellar objects (YSO) which create asymmetric outflows (jets) propagating up to the distances of a fraction of pc~\cite{mar93}. These jets may collide with dense molecular clouds forming bow shocks.
If a YSO is located within a molecular cloud (e.g. IRAS 16547-4247 or HH80), then hadrons, interacting with the distributed matter, could produce TeV $\gamma$-ray fluxes observable by CTA~\cite{ar07,bos10}.
The non-thermal radio emission, detected from the field of YSO BD+43◦3654~\cite{ben10}, indicate the importance  of non-thermal processes in these type of sources. A modelling of such emission predicts the production of TeV $\gamma$-rays above the sensitivity limit of CTA.

A transient acceleration of hadrons to TeV energies is also expected to occur in the magnetospheres of the accretion disks of low mass young stellar objects such as T Tauri stars~\cite{dv11}. These particles, colliding with matter, may produce $\gamma$-rays which, after absorption in the stellar and disk radiation, can escape only with energies below $\sim$100 GeV. Although, the expected $\gamma$-ray luminosities are relatively low, T Tauri stars can be promissing targets for the next generation Cherenkov telescopes. 
The transient nature of their $\gamma$-ray emission may however lower the detection probability. 
Note that there seems to exist a correlation between the $\gamma$-ray objects in the Fermi-LAT catalogue and those in the catalogue of known YSO~\cite{mun11}. 

Efforts to study massive young stellar objects and the T Tauri stars should be certainly undertaken with the sensitivity of CTA. These objects represent different conditions for possible acceleration of particles (e.g. low velocity jets but relatively dense medium). Their possible detection will provide interesting constraints on the acceleration processes of particles and formation of jets in conditions
significantly different from those expected in the sources with relativistic jets.

\subsection{Isolated White Dwarfs}

The processes for the high energy $\gamma$-ray production in the inner magnetospheres of rotating neutron stars (discussed in Sect.~2) may in principle operate also in the vicinity of the fast rotating White Dwarfs (WDs). Strongly magnetized, fast rotating WDs are expected to be a final stage of the evolution of the Ap and Bp type stars or as a result of mergers of WDs in the binary systems. In fact, strong magnetic fields have been discovered in the case of a few percent of isolated WDs and $\sim$25$\%$ of WDs in the Cataclysmic Variables~\cite{wf00}. 

The possibility that magnetized WDs can lose large amounts of the rotational energy on the generation of electromagnetic waves and acceleration of particles was considered in the context of two binary systems: 1E 2259+586~\cite{pacz90,us93} and AE Aquarii~\cite{dej94,ik98}. 
It is expected that magnetized WDs are able to inject relativistic electrons with energies allowing them to produce TeV $\gamma$-rays in the IC process (see e.g.~\cite{kas11}). In fact, a transient TeV emission from the binary system AE Aqr, containing fast rotating WD, was reported in the past~\cite{bow92,ch95}. These reports were not confirmed by more recent observations~\cite{lang98,si08}. However, due to its likely transient nature, the lack of confirmation does not exclude possible emission of $\gamma$-rays by such type of objects. Systematic observations of magnetized WDs with arrays of the next generation Cherenkov telescope might result in the detection of magnetospheric $\gamma$-ray emission from WDs. These $\gamma$-rays could be  produced as a result of ICS of thermal radiation from the hot stellar surfaces of the WDs. Note that due to the significantly weaker magnetic field in the inner magnetospheres of the rotating WDs, such pulsed $\gamma$-ray emission should extend to larger energies (GeV-TeV range) than those observed in $\gamma$-ray pulsars.
Moreover, steady TeV $\gamma$-ray emission, produced by leptons escaping from inner magnetospheres, could also appear as a result of the comptonization of the CMB or the infrared to optical photons produced in the Globular Clusters containing hundreds to thousands of magnetized WDs~\cite{bed12}. 

The sensitivity of CTA gives a chance to detect $\gamma$-rays from the nearby fast rotating magnetized WDs at distances up to $\sim 300$ pc (assuming typical period of 100 s, the surface magnetic field of $10^8$ G and the efficiency for $\gamma$-ray emission of 10$\%$)~\cite{bed12}.

\subsection{Accreting Neutron Stars (LMXBs)}

Old NSs in binary systems have usually lost most of their rotational energy.
Due to the long rotational periods, they are not able to create winds strong enough to push out the matter outside their light cylinder radii. They accrete matter from the companion star
in the two basic regimes, the so-called propeller and accretor stages. 
In the propeller stage, the compact object accretes matter only up to a certain distance from its surface, where the magnetic pressure of the rotating magnetosphere balances the pressure of the accreting matter. Such scenario was discussed in the past in the context of possible high energy processes in the so-called {\it hidden pulsar} model~\cite{ta91,tb93} and in the {\it cauldron} model~\cite{br84,tcl93}. More recently $\gamma$-ray emission from binary systems containing accreting NSs
was considered in~\cite{bed09a}. It was concluded that the turbulent transition region between the rotating magnetosphere of the compact object (propeller stage) and the accreting matter could provide good conditions for the acceleration of particles. In some systems, the particles might be able to produce even TeV $\gamma$-rays detectable by the Cherenkov telescopes. 

In the case of the accreting objects in compact Low Mass X-ray Binaries (LMXBs), huge gravitational energy of accreting matter is converted mainly into X-ray emission. 
It has been suspected that some of this power might also go to $\gamma$-rays. Therefore, LMXBs have been considered as potential targets for $\gamma$-ray telescopes. 
X-rays produced close to the accreting object or in the accretion disk can heat the nearby companion star to a temperature significantly above that obtained from nuclear burning. Therefore, it is likely that the irradiated stars produce soft radiation fields for relativistic  electrons comparable to those observed in the massive TeV $\gamma$-ray binaries (see~\cite{bp10}. 
Up to now, none of the LMXBs have been detected by the modern Cherenkov telescopes (see e.g.~\cite{ale11a}). However, the first object belonging to the class of compact Low Mass X-ray Binaries might be already detected in the GeV energies by the Fermi-LAT telescope~\cite{tam10}. These objects remain interesting targets for the future instruments with good sensitivity above a few tens GeV such as CTA.

\subsection{Binary systems containing MSPs}

A significant part of the millisecond pulsars (MSPs) are also found in the  compact binary systems. As discussed above for classical pulsars, also MSPs can produce winds which interact with winds of companion stars producing shocks able to accelerate electrons to GeV-TeV energies (e.g.~\cite{kl88,ph88,at93}.
The $\gamma$-ray fluxes from such binary systems are expected to be low due to much weaker radiation fields of the companion stars in these systems.
The soft radiation field from the surface of the companion star can be larger than that produced in nuclear burning in the case of a strong illumination of the stellar surface by X-rays from the pulsar. Then, the companion star may be heated to significantly larger  temperatures than those expected from the pure nuclear burning mechanism (e.g.~\cite{bp10}). Relativistic electrons from MSPs, interacting with this enhanced  stellar radiation, would produce $\gamma$-rays as a result of IC scattering. Detailed $\gamma$-ray light curves from such binaries are expected to differ significantly from those observed in the TeV $\gamma$-ray binaries due to the anisotropic emission of thermal radiation by the companion star. 

Some of the binary systems containing MSPs can be completely surrounded by the stellar winds creating the so called "hidden" MSPs~\cite{ta91}. In principle, such MSP binaries can also contribute to the $\gamma$-ray emission observed from individual objects or from Globular Clusters.

\section{Prospects for detection/identification of the Galactic $\gamma$-ray sources with CTA}

About a few hundred GeV $\gamma$-ray sources in the Galactic plane~\cite{abdo11a} and several TeV $\gamma$-ray sources (see TeVCat Catalogue and also~\cite{ah08b}) have no clear associations with known objects in the Galaxy.
Many of these sources observed in GeV energies are probably pulsars and in the TeV energies the PWNe surrounding them.
In fact, out of the 25 PWNe in the TeVCat Catalogue, 16 have associations with the
2nd Fermi Catalogue~\cite{abdo11a}. Many of the TeV unidentified sources show some angular extension what strengthen the hypothesis of their relation to PWNe or SNRs. On the other hand, out of the  unidentified TeV sources, 17 are associated with the 2nd Fermi Catalogue sources. 
The lack of clear objects, seen in radio or X-rays, in the directions of these TeV sources puts strong  
constraints on the $\gamma$-ray radiation mechanism (leptonic, hadronic ?), and also on the transport process of particles in the sources and on their evolution. Therefore, detailed investigation of this broad class of unidentified GeV-TeV $\gamma$-ray objects, with an order of magnitude more sensitive instrument and with the better angular resolution (CTA), will provide important constraints on the sites of particle acceleration in the Galaxy.

Significant amount of the unassociated GeV sources tend to group  in the directions to regions in the Galaxy with bright diffuse $\gamma$-ray emission. These sources often show curved spectra (about $28\%$~\cite{abdo11a}) suggesting that their TeV emission may be on a much lower level. One of the main tasks of CTA will be to provide information on the shape of the $\gamma$-ray spectra at sub-TeV energies of these sources in order to conclude on the spectral behaviour: Do they really cut-off or do they show a second high energy spectral component? For this purpose, the future instrument should have a low energy threshold in order to obtain good quality spectra at energies clearly below $\sim$100 GeV. The CTA superior sensitivity and good angular resolution should enable finding associations of these $\gamma$-ray sources with the objects known in other parts of electromagnetic spectrum or with already known classes of the $\gamma$-ray sources (e.g. SNRs, PWNe, binaries, etc ...).  

The large field of view of the CTA would make galactic plane scans possible with an order of magnitude better sensitivity than it has been done up to now by the HESS collaboration~\cite{ah05e,ah06f} and also in the case of the Cygnus region by the VERITAS Collaboration~\cite{al11b}. Such scans should be done
regularly in the same parts of the Galactic plane in order to have a chance to discover variable and periodic $\gamma$-ray sources. 
Up to now, a number of such variable known sources is relatively small probably due to the limited sensitivity of the present Cherenkov telescopes.
Frequent scanning of the same regions of the Galactic disk with CTA should result in discovery of  new classes of variable TeV $\gamma$-ray sources not observed up to now.

CTA should also have the power to investigate fine structures of the much extended sources of the type recently reported by the MILAGRO detector (e.g. MGRO J2019+37, MGRO J2031+41, C3, or C4~\cite{ab07a,ab07b}). An investigation of the specific parts of these sources (spectral information, possible variability ?) would allow us to conclude whether these TeV sources are due to either the composition of a few nearby different types of sources  (e.g. in the Cygnus region) or they represent separate parts of a nearby source (e.g. source C3 coincident with Geminga pulsar?). With the present ARGO-YBJ and the planned construction of the next generation of the water tank Cherenkov detectors (HAWC and LHAASO), the number of such extended sources will certainly increase being a very interesting class of objects to be studied with CTA.

\section{Conclusion}

The important role of the high energy processes in a few types of compact Galactic sources have been recently undoubtedly established with the discovery of a few hundred GeV and a few tens of TeV $\gamma$-ray sources. Many surprising emission features have been already discovered. Their detailed investigation will certainly put a new light on the origin of the high energy particles in the Galaxy. The nature of the high energy processes resulting in 
the production of the high energy $\gamma$-rays remains mostly unknown and is at present debated. The appearance of the $\gamma$-ray emission in the case of some sources have not been expected by theoretical modelling. In order to understand these high energy phenomena, extensive deep studies with the next generation of the TeV telescopes are required. In fact, the scientific community realized these needs initiating new instrumental projects in the high energy $\gamma$-ray astronomy
such as CTA, HAWC, or LHAASO. They will be able to study the already discovered classes of objects with an order of magnitude better sensitivity. The main tasks of the future TeV observations will be:

\begin{itemize}

\item Enlarge the number of sources in specific classes of $\gamma$-ray objects in order to enable comparative studies.

\item Discover new types of sources. The importance of high energy processes, leading to the $\gamma$-ray production, is still expected 
in specific objects based on their theoretical modelling.

\item Perform multiwavelength observations of the known types of sources (mainly in radio, X-ray, GeV and TeV $\gamma$-rays)
in order to get complementary information on their high energy features.

\item Perform morphological studies of the extended sources through the sub-TeV ($\sim$ 30 GeV) to multi-TeV ($>$30 TeV)
energies in order to determine the sites of particle acceleration, their propagation conditions within the sources and their surroundings, and answer the question whether the specific parts of the extended sources have common origin (single source of relativistic particles and/or radiation mechanism) or appear by chance as a result of overlapping of different (types of) sources. For example, in the case of the open clusters: pulsars, supernova remnants, binary systems and microquasars can contribute simultaneously to the observed $\gamma$-ray flux.

\item Try to determine the highest energy end in the TeV $\gamma$-ray spectra of the sources in order to get constraints on the maximum energies of accelerated particles and on the acceleration process.

\item Establish the shortest possible variability time scale of the TeV $\gamma$-ray emission in known (and new) types of sources in order to obtain constraints on the acceleration and cooling time scales of relativistic particles and perhaps answer the question on their production mechanism (hadronic or leptonic ?).

\item Study the spectral properties of the periodic $\gamma$-ray sources in order to investigate the acceleration of particles
and their radiation processes at different conditions at the source (magnetic field strength, density of matter and radiation fields).

\item Study the long term variability of periodic sources 
(pulsars, binary systems) and try to link these features to the conditions at the source derived from other observations (e.g. 
variability in the radio light curve of pulsars or their giant flares, or stellar cycles (similar to Solar cycles) resulting in the change of stellar wind properties?).

\item Establish a relation between the $\gamma$-ray sources 
observed in the GeV energy range (by Fermi-LAT and AGILE) to their properties in the TeV energy range (investigate spectral and morphological features). This should make possible to put constraints on the radiation processes engaged in the specific sources (importance of the absorption at the source and cascading processes?) and determine the main production process (hadronic versus leptonic?).

\item Study a relation of the variable $\gamma$-ray components 
in the sources (produced by particles accelerated recently) to the possible lower level steady components (produced in the source or around it by particles accelerated in the past) in order to constrain evolutionary properties of these sources.

\end{itemize}

The outcome of the investigations summarized above will provide 
conclusive information on the physics of sources discussed here (such as the particle acceleration processes, properties of relativistic magnetized plasma, structure of rotating magnetospheres, formation of relativistic jets, propagation of particles, etc ...).   
It is hoped that, with the next generation of instruments, the basic question on the origin of energetic particles in the Galaxy will be finally answered. At least the author of this paper is convinced that this will happen.

\section*{Acknowledgment}
I would like to thank the reviewers for comments and many suggestions
which allowed me to improve the style of the paper. I also thank Maria Giller for reading the manuscript and useful comments. This work is supported by the Polish grants from Narodowe Centrum Bada\'n i Rozwoju "ERA-NET-ASPERA/01/10" and from Naro\-do\-we Cen\-trum Nau\-ki "2011/01/M/ST9/01891".






\end{document}